\documentclass[runningheads]{llncs}
\usepackage{graphicx}

\usepackage{tikz}
\usepackage{comment}
\usepackage{amsmath,amssymb} 
\usepackage{color}

\newsavebox\CBox
\def\textBF#1{\sbox\CBox{#1}\resizebox{\wd\CBox}{\ht\CBox}{\textbf{#1}}}

\newcommand{\eg}{{\emph{e.g.}}}
\newcommand{\ie}{{\emph{i.e.}}}

\newcommand{\etc}{{\emph{etc}}}
\definecolor{grayhighlight}{RGB}{213,229,255}

\begin{document}
\pagestyle{headings}
\mainmatter
\def\ECCVSubNumber{100}  

\title{Efficient and Accurate Quantized Image Super-Resolution on Mobile NPUs, Mobile AI \& AIM 2022 challenge: Report} 

\titlerunning{Efficient and Accurate Quantized Image Super-Resolution on Mobile NPUs}
\authorrunning{ECCV-22 submission ID \ECCVSubNumber}
\author{Anonymous ECCV submission}
\institute{Paper ID \ECCVSubNumber}

\author{Andrey Ignatov \and Radu Timofte \and Maurizio Denna \and Abdel Younes \and
Ganzorig Gankhuyag \and Jingang Huh \and Myeong Kyun Kim \and Kihwan Yoon \and Hyeon-Cheol Moon \and Seungho Lee \and Yoonsik Choe \and Jinwoo Jeong \and Sungjei Kim \and
Maciej Smyl \and Tomasz Latkowski \and Pawel Kubik \and Michal Sokolski \and Yujie Ma \and 
Jiahao Chao \and Zhou Zhou \and Hongfan Gao \and Zhengfeng Yang \and Zhenbing Zeng \and
Zhengyang Zhuge \and Chenghua Li \and
Dan Zhu \and Mengdi Sun \and Ran Duan \and Yan Gao \and
Lingshun Kong \and Long Sun \and Xiang Li \and Xingdong Zhang \and Jiawei Zhang \and Yaqi Wu \and Jinshan Pan \and
Gaocheng Yu \and Jin Zhang \and Feng Zhang \and Zhe Ma \and Hongbin Wang \and
Hojin Cho \and Steve Kim \and
Huaen Li \and Yanbo Ma \and
Ziwei Luo \and Youwei Li \and Lei Yu \and Zhihong Wen \and Qi Wu \and Haoqiang Fan \and Shuaicheng Liu \and
Lize Zhang \and Zhikai Zong \and
Jeremy Kwon \and
Junxi Zhang \and Mengyuan Li \and Nianxiang Fu \and Guanchen Ding \and Han Zhu \and Zhenzhong Chen \and
Gen Li \and Yuanfan Zhang \and  Lei Sun \and 
Dafeng Zhang \and
Neo Yang \and Fitz Liu \and Jerry Zhao \and
Mustafa Ayazoglu \and Bahri Batuhan Bilecen \and
Shota Hirose \and Kasidis Arunruangsirilert \and Luo Ao \and
Ho Chun Leung \and Andrew Wei \and Jie Liu \and Qiang Liu \and Dahai Yu \and
Ao Li \and Lei Luo \and Ce Zhu \and
Seongmin Hong  \and Dongwon Park \and Joonhee Lee \and Byeong Hyun Lee \and Seunggyu Lee \and Se Young Chun \and
Ruiyuan He \and Xuhao Jiang \and
Haihang Ruan \and Xinjian Zhang \and Jing Liu \and
Garas Gendy \and Nabil Sabor \and Jingchao Hou \and Guanghui He  $^*$
}

\institute{}
\authorrunning{A. Ignatov, R. Timofte, M. Denna, A. Younes et al.}
\maketitle

\begin{abstract}

Image super-resolution is a common task on mobile and IoT devices, where one often needs to upscale and enhance low-resolution images and video frames. While numerous solutions have been proposed for this problem in the past, they are usually not compatible with low-power mobile NPUs having many computational and memory constraints. In this Mobile AI challenge, we address this problem and propose the participants to design an efficient quantized image super-resolution solution that can demonstrate a real-time performance on mobile NPUs. The participants were provided with the DIV2K dataset and trained INT8 models to do a high-quality 3X image upscaling. The runtime of all models was evaluated on the Synaptics VS680 Smart Home board with a dedicated edge NPU capable of accelerating quantized neural networks.  All proposed solutions are fully compatible with the above NPU, demonstrating an up to 60 FPS rate when reconstructing Full HD resolution images. A detailed description of all models developed in the challenge is provided in this paper.

\end{abstract}

{\let\thefootnote\relax\footnotetext{%
$^*$ Andrey Ignatov \textit{(andrey@vision.ee.ethz.ch)}, Radu Timofte \textit{(radu.timofte@uni-wuerzburg.de)}, Maurizio Denna \textit{(maurizio.denna@synaptics.com)} and Abdel Younes (\textit{abdel.younes@synaptics.com}) are the Mobile AI \& AIM 2022 challenge organizers. The other authors participated in the challenge. \\ Appendix \ref{sec:apd:team} contains the authors' team names and affiliations. \vspace{2mm} \\ Mobile AI 2022 Workshop website: \\ \url{https://ai-benchmark.com/workshops/mai/2022/}
}}

\section{Introduction}

\begin{figure*}[t!]
\centering
\setlength{\tabcolsep}{1pt}
\resizebox{\linewidth}{!}
{
\includegraphics[width=0.5\linewidth]{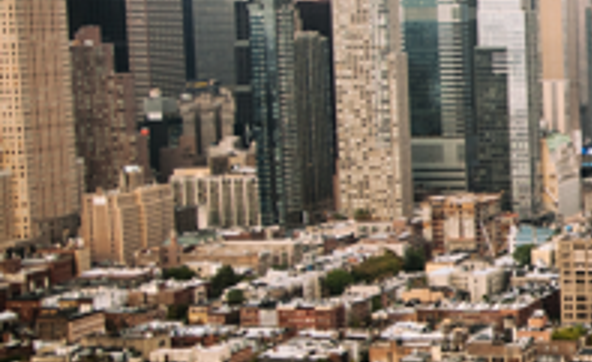} \hspace{2mm}
\includegraphics[width=0.5\linewidth]{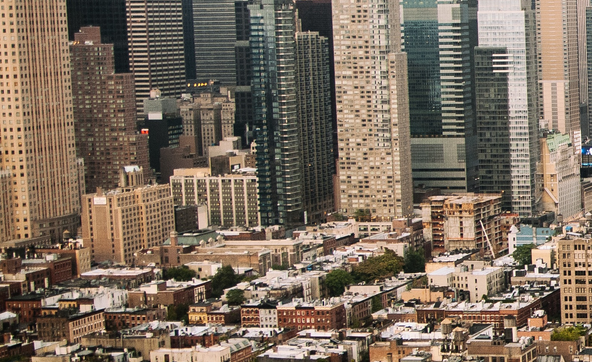}
}
\vspace{0cm}
\caption{Sample crop from a 3X bicubically upscaled image and the target DIV2K~\cite{agustsson2017ntire} photo.}
\label{fig:example_photos}
\end{figure*}

Single image super-resolution is a longstanding computer vision problem. Its goal is to restore the original image contents from its downsampled version by recovering lost details.
This task draws ever-increasing interest due to the pervasive media contents, cameras and displays and its direct application to real-world problems such as image processing in smartphone cameras (\eg telephoto), enhancement of low-resolution media data and upscaling images and videos to match the high resolution of the displays. There is a rich literature and a multitude of classical (hand-crafted)~\cite{irani1991improving,freeman2002example,park2003super,timofte2013anchored,timofte2014a+,yang2013fast,yang2008image,yang2010image,huang2015single,timofte2016seven} and deep learning-based~\cite{dong2015image,dong2014learning,kim2016accurate,lim2017enhanced,timofte2017ntire,timofte2018ntire,cai2019ntire,lugmayr2020ntire,zhang2020ntire,ignatov2018pirm} methods have been suggested in the past decades. One of the biggest drawbacks of the existing solutions is that the were not optimized for computational efficiency or mobile-related constraints, but for sheer accuracy in terms of high fidelity scores.
Meeting the hardware-constraints from low-power devices is essential for the methods developed for real-world applications of image super-resolution and other tasks related to image processing and enhancement~\cite{ignatov2017dslr,ignatov2018wespe,ignatov2020replacing} on mobile devices. In this challenge, we address this drawback and by using the popular DIV2K~\cite{agustsson2017ntire} image super-resolution dataset, we add efficiency-related constraints from mobile NPUs on the images super-resolution solutions.


The deployment of AI-based solutions on portable devices usually requires an efficient model design based on a good understanding of the mobile processing units (\eg CPUs, NPUs, GPUs, DSP) and their hardware particularities, including their memory constraints. We refer to~\cite{ignatov2019ai,ignatov2018ai} for an extensive overview of mobile AI acceleration hardware, its particularities and performance. As shown in these works, the latest generations of mobile NPUs are reaching the performance of older-generation mid-range desktop GPUs. Nevertheless, a straightforward deployment of neural networks-based solutions on mobile devices is impeded by (i) a limited memory (\ie, restricted amount of RAM) and
(ii) a limited or lacking support of many common deep learning operators and layers. These impeding factors make the processing of high resolution inputs impossible with the standard NN models and require a careful adaptation or re-design to the constraints of mobile AI hardware. Such optimizations can employ a combination of various model techniques such as 16-bit / 8-bit~\cite{chiang2020deploying,jain2019trained,jacob2018quantization,yang2019quantization} and low-bit~\cite{cai2020zeroq,uhlich2019mixed,ignatov2020controlling,liu2018bi} quantization, network pruning and compression~\cite{chiang2020deploying,ignatov2020rendering,li2019learning,liu2019metapruning,obukhov2020t}, device- or NPU-specific adaptations, platform-aware neural architecture search~\cite{howard2019searching,tan2019mnasnet,wu2019fbnet,wan2020fbnetv2}, \etc.

The majority of competitions aimed at efficient deep learning models use standard desktop hardware for evaluating the solutions, thus the obtained models rarely show acceptable results when running on real mobile hardware with many specific constraints. In this \textit{Mobile AI challenge}, we take a radically different approach and propose the participants to develop and evaluate their models directly on mobile devices. The goal of this competition is to design a fast and performant quantized deep learning-based solution for image super-resolution problem. For this, the participants were provided with the large-scale DIV2K~\cite{agustsson2017ntire} dataset containing diverse 2K resolution RGB images used to train their models using a downscaling factor of 3. The efficiency of the proposed solutions was evaluated on the Synaptics Dolphin platform featuring a dedicated NPU that can efficiently accelerate INT8 neural networks. The overall score of each submission was computed based on its fidelity and runtime results, thus balancing between the image reconstruction quality and the efficiency of the model. All solutions developed in this challenge are fully compatible with the TensorFlow Lite framework~\cite{TensorFlowLite2021}, thus can be executed on various Linux and Android-based IoT platforms, smartphones and edge devices.

This challenge is a part of the \textit{Mobile AI \& AIM 2022 Workshops and Challenges} consisting of the following competitions:

\small

\begin{itemize}
\item Quantized Image Super-Resolution on Mobile NPUs
\item Power Efficient Video Super-Resolution on Mobile NPUs~\cite{ignatov2022maivideosr}
\item Learned Smartphone ISP on Mobile GPUs~\cite{ignatov2022maiisp}
\item Efficient Single-Image Depth Estimation on Mobile Devices~\cite{ignatov2022maidepth}
\item Realistic Bokeh Effect Rendering on Mobile GPUs~\cite{ignatov2022maibokeh}
\item Super-Resolution of Compressed Image and Video~\cite{yang2022aim}
\item Reversed Image Signal Processing and RAW Reconstruction~\cite{conde2022aim}
\item Instagram Filter Removal~\cite{kinli2022aim}
\end{itemize}

\noindent The results and solutions obtained in the previous \textit{MAI 2021 Challenges} are described in our last year papers:

\small

\begin{itemize}
\item Learned Smartphone ISP on Mobile NPUs~\cite{ignatov2021learned}
\item Real Image Denoising on Mobile GPUs~\cite{ignatov2021fastDenoising}
\item Quantized Image Super-Resolution on Mobile NPUs~\cite{ignatov2021real}
\item Real-Time Video Super-Resolution on Mobile GPUs~\cite{romero2021real}
\item Single-Image Depth Estimation on Mobile Devices~\cite{ignatov2021fastDepth}
\item Quantized Camera Scene Detection on Smartphones~\cite{ignatov2021fastSceneDetection}
\end{itemize}

\normalsize


\begin{figure*}[t!]
\centering
\setlength{\tabcolsep}{1pt}
\resizebox{0.96\linewidth}{!}
{
\includegraphics[width=1.0\linewidth]{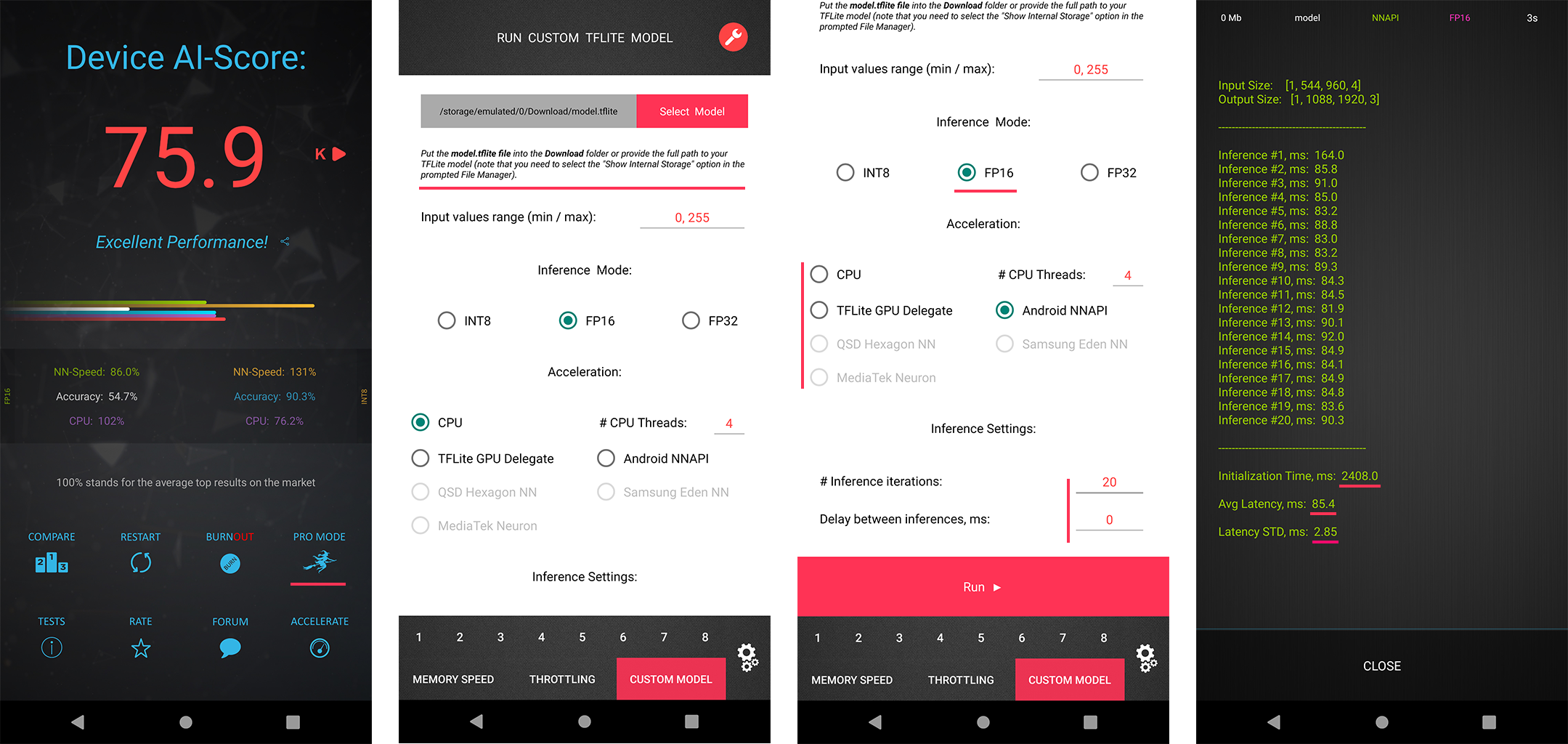}
}
\vspace{0.2cm}
\caption{Loading and running custom TensorFlow Lite models with AI Benchmark application. The currently supported acceleration options include Android NNAPI, TFLite GPU, Hexagon NN, Qualcomm QNN, MediaTek Neuron and Samsung ENN delegates as well as CPU inference through TFLite or XNNPACK backends. The latest app version can be downloaded at \url{https://ai-benchmark.com/download}}
\label{fig:ai_benchmark_custom}
\end{figure*}

\section{Challenge}

In order to design an efficient and practical deep learning-based solution for the considered task that runs fast on mobile devices, one needs the following tools:

\begin{enumerate}
\item A large-scale high-quality dataset for training and evaluating the models. Real, not synthetically generated data should be used to ensure a high quality of the obtained model;
\item An efficient way to check the runtime and debug the model locally without any constraints as well as the ability to check the runtime on the target evaluation platform.
\end{enumerate}

This challenge addresses all the above issues. Real training data, tools, and runtime evaluation options provided to the challenge participants are described in the next sections.

\subsection{Dataset}

In this challenge, the participants were proposed to work with the popular DIV2K~\cite{agustsson2017ntire} dataset. It consists from 1000 divers 2K resolution RGB images: 800 are used for training, 100 for validation and 100  for testing purposes. The images are of high quality both aesthetically and in the terms of small amounts of noise and other corruptions (like blur and color shifts). All images were manually collected and have 2K pixels on at least one of the axes (vertical or horizontal). DIV2K covers a large diversity of contents, from people, handmade objects and environments (cities), to  flora and fauna and natural sceneries, including underwater. An example set of images is demonstrated in Fig.~\ref{fig:example_photos}.

\subsection{Local Runtime Evaluation}

When developing AI solutions for mobile devices, it is vital to be able to test the designed models and debug all emerging issues locally on available devices. For this, the participants were provided with the \textit{AI Benchmark} application~\cite{ignatov2018ai,ignatov2019ai} that allows to load any custom TensorFlow Lite model and run it on any Android device with all supported acceleration options. This tool contains the latest versions of \textit{Android NNAPI, TFLite GPU, Hexagon NN, Qualcomm QNN, MediaTek Neuron} and \textit{Samsung ENN} delegates, therefore supporting all current mobile platforms and providing the users with the ability to execute neural networks on smartphone NPUs, APUs, DSPs, GPUs and CPUs.

\smallskip

To load and run a custom TensorFlow Lite model, one needs to follow the next steps:

\begin{enumerate}
\setlength\itemsep{0mm}
\item Download AI Benchmark from the official website\footnote{\url{https://ai-benchmark.com/download}} or from the Google Play\footnote{\url{https://play.google.com/store/apps/details?id=org.benchmark.demo}} and run its standard tests.
\item After the end of the tests, enter the \textit{PRO Mode} and select the \textit{Custom Model} tab there.
\item Rename the exported TFLite model to \textit{model.tflite} and put it into the \textit{Download} folder of the device.
\item Select mode type \textit{(INT8, FP16, or FP32)}, the desired acceleration/inference options and run the model.
\end{enumerate}

\noindent These steps are also illustrated in Fig.~\ref{fig:ai_benchmark_custom}.

\subsection{Runtime Evaluation on the Target Platform}

In this challenge, we use the \textit{Synaptics VS680 Edge AI SoC} \cite{SynapticsSoC2021} Evaluation Kit as our target runtime evaluation platform. The VS680 Edge AI SoC is integrated into Smart Home solution and it features a powerful NPU designed by \textit{VeriSilicon} and capable of accelerating quantized models (up to 7 TOPS). It supports Android and can perform NN inference through NNAPI, demonstrating INT8 AI Benchmark scores that are close to the ones of mid-range smartphone chipsets. Within the challenge, the participants were able to upload their TFLite models to an external server and get feedback regarding the speed of their model: the inference time of their solution on the above mentioned NPU or an error log if the network contained incompatible operations and/or improper quantization. Participants' models were first compiled and optimized using Synaptics' SyNAP Toolkit and then executed on the the VS680's NPU using the SyNAP C++ API to achieve the best possible efficiency. The same setup was also used for the final runtime evaluation. The participants were additionally provided with a list of ops supported by this board and model optimization guidance in order to fully utilize the NPU's convolution and tensor processing resources. Besides that, a layer-by-layer timing information was provided for each submitted TFLite model to help the participants to optimize the architecture of their networks.

\subsection{Challenge Phases}

The challenge consisted of the following phases:

\vspace{-0.8mm}
\begin{enumerate}
\item[I.] \textit{Development:} the participants get access to the data and AI Benchmark app, and are able to train the models and evaluate their runtime locally;
\item[II.] \textit{Validation:} the participants can upload their models to the remote server to check the fidelity scores on the validation dataset, to get the runtime on the target platform, and to compare their results on the validation leaderboard;
\item[III.] \textit{Testing:} the participants submit their final results, codes, TensorFlow Lite models, and factsheets.
\end{enumerate}
\vspace{-0.8mm}

\subsection{Scoring System}

All solutions were evaluated using the following metrics:

\vspace{-0.8mm}
\begin{itemize}
\setlength\itemsep{-0.2mm}
\item Peak Signal-to-Noise Ratio (PSNR) measuring fidelity score,
\item Structural Similarity Index Measure (SSIM), a proxy for perceptual score,
\item The runtime on the target Synaptics VS680 board.
\end{itemize}
\vspace{-0.8mm}

The score of each final submission was evaluated based on the next formula ($C$ is a constant normalization factor):

\smallskip
\begin{equation*}
\text{Final Score} \,=\, \frac{2^{2 \cdot \text{PSNR}}}{C \cdot \text{runtime}},
\end{equation*}
\smallskip

During the final challenge phase, the participants did not have access to the test dataset. Instead, they had to submit their final TensorFlow Lite models that were subsequently used by the challenge organizers to check both the runtime and the fidelity results of each submission under identical conditions. This approach solved all the issues related to model overfitting, reproducibility of the results, and consistency of the obtained runtime/accuracy values.

\section{Challenge Results}

\begin{table*}[t!]
\centering
\resizebox{\linewidth}{!}
{
\begin{tabular}{l|c|cc|ccc|ccc|c}
\hline
Team \, & \, Author \, & \, Framework \, & \, Model Size, \, & \, PSNR$\uparrow$ \, & \, SSIM$\uparrow$ \, & \, $\Delta$ PSNR Drop, \, & \multicolumn{2}{c}{\, Runtime, ms $\downarrow$ \,} & Speed-Up & \, Final Score \\
& & & \small{KB} & \multicolumn{2}{c}{\small{INT8 Model}} & \small{FP32 $\to$ INT8} & \, \small{CPU} \, & \, \small{NPU} \, & & \\
\hline
\hline
Z6 & ganzoo & Keras / TensorFlow & 67 & 30.03 & 0.8738 & 0.06 & 809 & 19.2 & 42.1 & 22.22 \\
TCLResearchEurope \, & maciejos\_s & \, Keras / TensorFlow \, & 53 & 29.88 & 0.8705 & 0.13 & 824 & 15.9 & 51.8 & 21.84 \\
ECNUSR & CCjiahao & Keras /TensorFlow & 50 & 29.82 & 0.8697 & 0.10 & 553 & 15.1 & 36.6 & 21.08 \\
LCVG & zion & Keras / TensorFlow & 48 & 29.76 & 0.8675 & 0.11 & 787	 & \textBF{15.0} & 52.5 & 19.59 \\
BOE-IOT-AIBD & NBCS & Keras / TensorFlow & 57 & 29.80 & 0.8675 & 0.19 & 877 & 16.1 & 54.5 & 19.27 \\
NJUST & kkkls & TensorFlow & 57 & 29.76 & 0.8676 & 0.16 & 767 & 15.8 & 48.5 & 18.56 \\
Antins\_cv & fz & Keras / TensorFlow & 38 & 29.58 & 0.8609 & n.a. & \textBF{543} & 15.2 & 35.7 & 15.02 \\
GenMedia Group & stevek & Keras / TensorFlow & 56 & 29.90 & 0.8704 & 0.11 & 855 & 25.6 & 33.4 & 13.91 \\
Vccip & Huaen & \, PyTorch / TensorFlow \, & 67 & 29.98 & 0.8729 & n.a. & 1042 & 30.5 & 34.2 & 13.07 \\
MegSR & balabala & Keras / TensorFlow & 65 & 29.94 & 0.8704 & n.a. & 965 & 29.8 & 32.4 & 12.65 \\
DoubleZ & gideon & Keras / TensorFlow & 63 & 29.94 & 0.8712 & 0.1 & 990 & 30.1 & 32.9 & 12.54 \\
Jeremy Kwon & alan\_jaeger & TensorFlow & 48 & 29.80 & 0.8691 & 0.09 & 782 & 25.7 & 30.4 & 12.09 \\
Lab216 & sissie & TensorFlow & 78 & 29.94 & 0.8728 & 0 & 1110 & 31.8 & 34.9 & 11.85 \\
TOVB & jklovezhang & Keras / TensorFlow & 56 & 30.01 & 0.8740 & 0.04 & 950 & 43.3 & 21.9 & 9.60 \\
\hline
ABPN~\cite{du2021anchor} & baseline & - & 53 & 29.87 & 0.8686 & n.a. & 998 & 36.9 & 27 & 9.27 \\
\hline
Samsung Research & xiaozhazha & TensorFlow & 57 & 29.95 & 0.8728 & 0.1 & 941 & 43.2 & 21.8 & 8.84 \\
Rtsisr2022 & rtsisr & Keras / TensorFlow & 68 & 30.00 & 0.8729 & 0.09 & 977 & 46.4 & 21.1 & 8.83 \\
Aselsan Research & deepernewbie & Keras / TensorFlow & 30 & 29.67 & 0.8651 & n.a. & 598 & 30.2 & 19.8 & 8.59 \\
Klab\_SR & FykAikawa & TensorFlow & 39 & 29.88 & 0.8700 & n.a. & 850 & 43.1 & 19.7 & 8.05 \\
TCL Research HK &  mrblue & Keras / TensorFlow & 121 & \textBF{30.10} & \textBF{0.8751} & 0.04 & 1772 & 60.2 & 29.4 & 7.81 \\
RepGSR & \, yilitiaotiaotang \, & Keras / TensorFlow & 90 & 30.06 & 0.8739 & 0.02 & 1679 & 61.3 & 27.4 & 7.26 \\
ICL & smhong & Keras / TensorFlow & 55 & 29.76 & 0.8641 & 0.23 & 949 & 43.2 & 22.0 & 6.79 \\
Just A try & kia350 & Keras / TensorFlow & 39 & 29.75 & 0.8671 & n.a. & 766 & 42.9 & 17.9 & 6.76 \\
Bilibili AI & Sail & Keras / TensorFlow & 75 & 29.99 & 0.8729 & n.a. & 1075 & 68.2 & 15.8 & 5.92 \\
MobileSR &  garasgaras & TensorFlow & 82 & 30.02 & 0.8735 & 0.07 & 1057 & 72.4 & 14.6 & 5.82 \\
\hline
A+ regression~\cite{timofte2014a+} & Baseline &  &  & 29.32 & 0.8520 & - & - & - & -  & - \\
Bicubic Upscaling & Baseline &  &  & 28.26 & 0.8277 & - & - & - & - & - \\
\hline
\end{tabular}
}
\vspace{2.6mm}
\caption{\small{Mobile AI 2022 Real-Time Image Super-Resolution challenge results and final rankings. During the runtime measurements, the models were performing image upscaling from 640$\times$360 to 1920$\times$1080 pixels. $\Delta$ PSNR values correspond to accuracy loss measured in comparison to the original floating-point network. Team \textit{Z6} is the challenge winner.}}
\label{tab:results}
\end{table*}

From above 250 registered participants, 28 teams entered the final phase and submitted their results, TFLite models, codes, executables and factsheets. Table~\ref{tab:results} summarizes the final challenge results and reports PSNR, SSIM and runtime numbers for the valid solutions on the final test dataset and on the target evaluation platform. The proposed methods are described in section~\ref{sec:solutions}, and the team members and affiliations are listed in Appendix~\ref{sec:apd:team}.

\subsection{Results and Discussion}

All solutions proposed in this challenge demonstrated an extremely high efficiency, being able to reconstruct a Full HD image on the target Synaptics VS680 board under 15-75 ms, significantly surpassing the basic bicubic image upsampling baseline in terms of the resulting image quality. Moreover, 14 teams managed to beat the last year's top solution~\cite{du2021anchor}, demonstrating a better runtime and/or accuracy results. This year, all teams performed INT8 model quantization correctly, and the accuracy drop caused by FP32 $\to$ INT8 conversion is less than 0.1 dB in the majority of cases. The general recipe of how to get a fast quantized image super-resolution model compatible with the latest mobile NPUs is as follows:
\begin{enumerate}
\item Use a shallow network architecture with or without skip connections;
\item Limit the maximum convolution filter size to $3\times3$;
\item Use the depth-to-space layer for the final image upsampling instead of the transposed convolution;
\item Use re-parameterized convolution blocks that are fused into one single convolution layer during the inference for better fidelity results;
\item Replace the nearest neighbor upsampling op with an equivalent 1$\times$1 convolution layer for a better latency;
\item Clip the model's output values to avoid incorrect output normalization during INT8 quantization; place the clip op before the depth-to-space layer;
\item Fine-tune the network with the quantization aware training algorithm before performing the final INT8 model conversion;
\item Transfer learning might also be helpful in some cases.
\end{enumerate}
Team \textit{Z6} is the challenge winner~--- the proposed solution demonstrated over 50 FPS on the Synaptics platform while also achieved one of the best fidelity results. Despite the simplicity of the presented SCSRN model, the authors used an advanced training strategy with weights clipping and normalization to get good image reconstruction results. The second best solution was obtained by team \textit{TCLResearchEurope}, which model achieved more than 62 FPS on the Synaptics VS680 board, thus allowing for a 1080P@60FPS video upscaling. The authors used a combination of the convolution re-parametrization, channel pruning and model distillation techniques to get a small but powerful image super-resolution network. The best numerical results in this competition were achieved by team \textit{TCL Research HK} due to a deep structure, convolution re-parametrization, and a large number of skip connections in the proposed model architecture.

When it comes to the Synaptics VS680 smart TV board itself, we can also see a noticeable improvement in its performance compared to the last year's results, which is caused by a number of optimizations in its NPU and NNAPI drivers. In particular, this year it was able to efficiently accelerate all models submitted during the final phase of the competition, showing a speed up of up to 20-55 times compared to CPU-based model execution. This demonstrates that an accurate deep learning-based high-resolution video upsampling on IoT platforms is a reality rather than just a pure concept, and the solutions proposed in this challenge are already capable to reconstruct Full HD videos on this SoC. With additional optimizations, it should also be possible to perform HD to 4K video upsampling, which is nowadays a critical task for many smart TV boards.

\section{Challenge Methods}
\label{sec:solutions}

\noindent This section describes solutions submitted by all teams participating in the final stage of the MAI 2022 Real-Time Image Super-Resolution challenge.

\subsection{Z6}

\begin{figure*}[h!]
\centering
\includegraphics[width=0.7\linewidth]{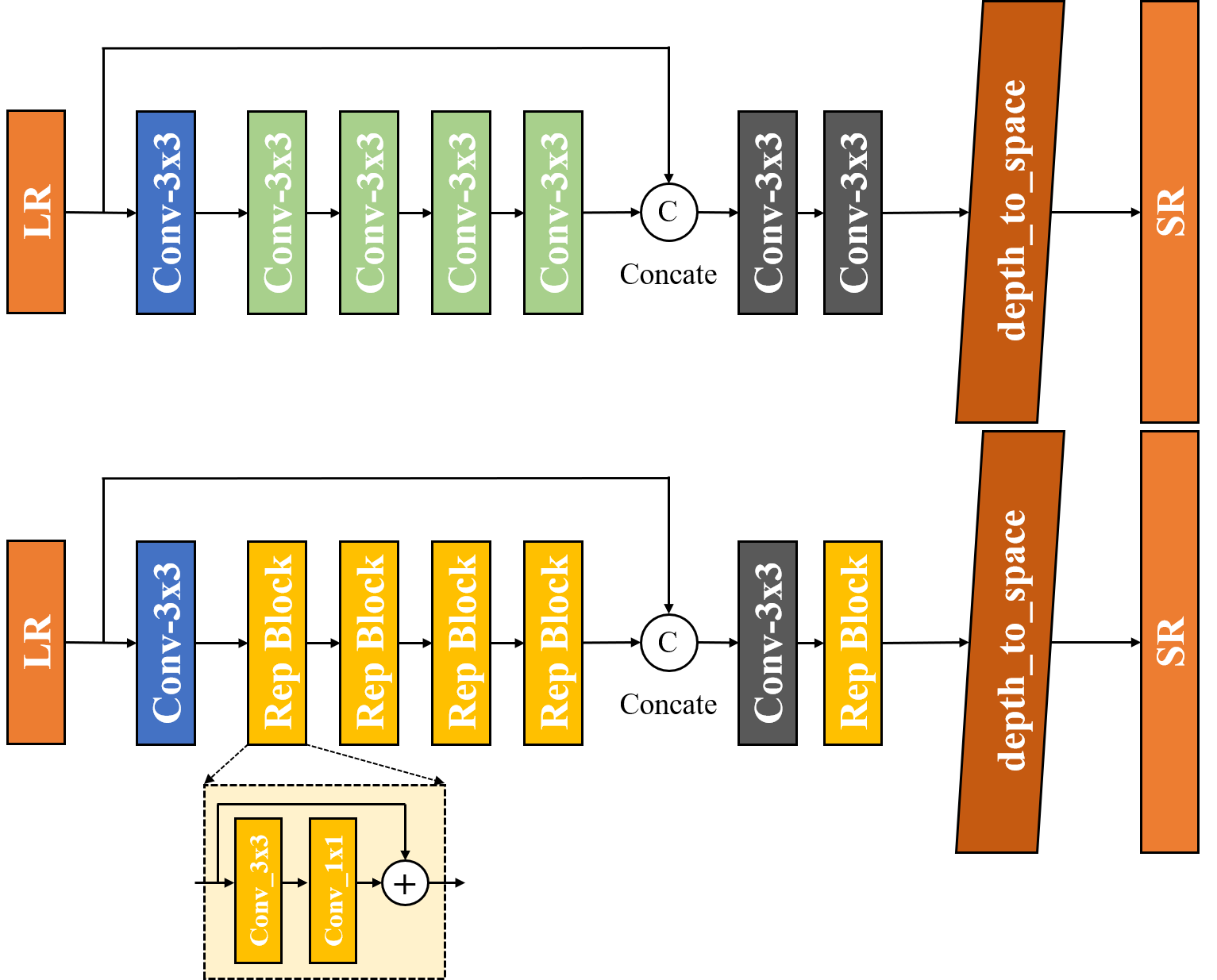}
\caption{\small{SCSRN model architecture proposed by team Z6 (Top) inference network (Bottom) train network.}}
\label{fig:Z6}
\end{figure*}

Team Z6 proposed a compact Skip-Concatenated Image Super Resolution Network (SCSRN) for the considered task (Fig.~\ref{fig:Z6}). This model consists of twoseven $3\times3$ convolution layers, five Re-Parameterizable blocks (Rep\_Block) and a skip connection (concatenation of input image and intermediate feature map directly). The number of channels in the network is set to 32, the depth-to-space op is used to produce the final image output. The authors applied the weight clipping method during the training stage in order to mitigate the performance degradation after INT8 quantization: the weight distribution analysis revealed that the distribution is heavily skewed (asymmetric), especially in the first convolutional layer. In TFLite, the weights quantization is only performed in a symmetric way, thus the quantization error is significantly accumulated from the first layer. Additionally, the inference time was reduced and the correct model output was ensured by clipping the output values before the depth-to-space layer. The network was trained in three stages:

\begin{enumerate}
\item In the first stage, the model was trained from scratch using the L1 loss function on patches of size $128\times128$ pixels with a mini-batch size of 16. Network parameters were optimized for 800 epochs using the Adam optimizeralgorithm with a learning rate of 1e--3 decreased with the cosine warm-up scheduler~\cite{loshchilov2016sgdr} with a 0.1 percentage warm-up ratio.

\item Next, the obtained model was further trained to minimize the L2 loss for 200 epochs. The initial learning rate was set to 2e--5 and halved every 60 epochs. At this stage, we also apply a channel shuffle augmentation.

\item In the third stage, the model was fine-tuned with the quantization aware training algorithm for 300 epochs. The initial learning rate was set to 1e--54 and halved every 60 epochs. The authors used the DCT (Discrete Cosine Transform)-domain L1 loss function as the target metric.
\end{enumerate}

\subsection{TCLResearchEurope}

\begin{figure*}[h!]
\centering
\includegraphics[width=0.7\linewidth]{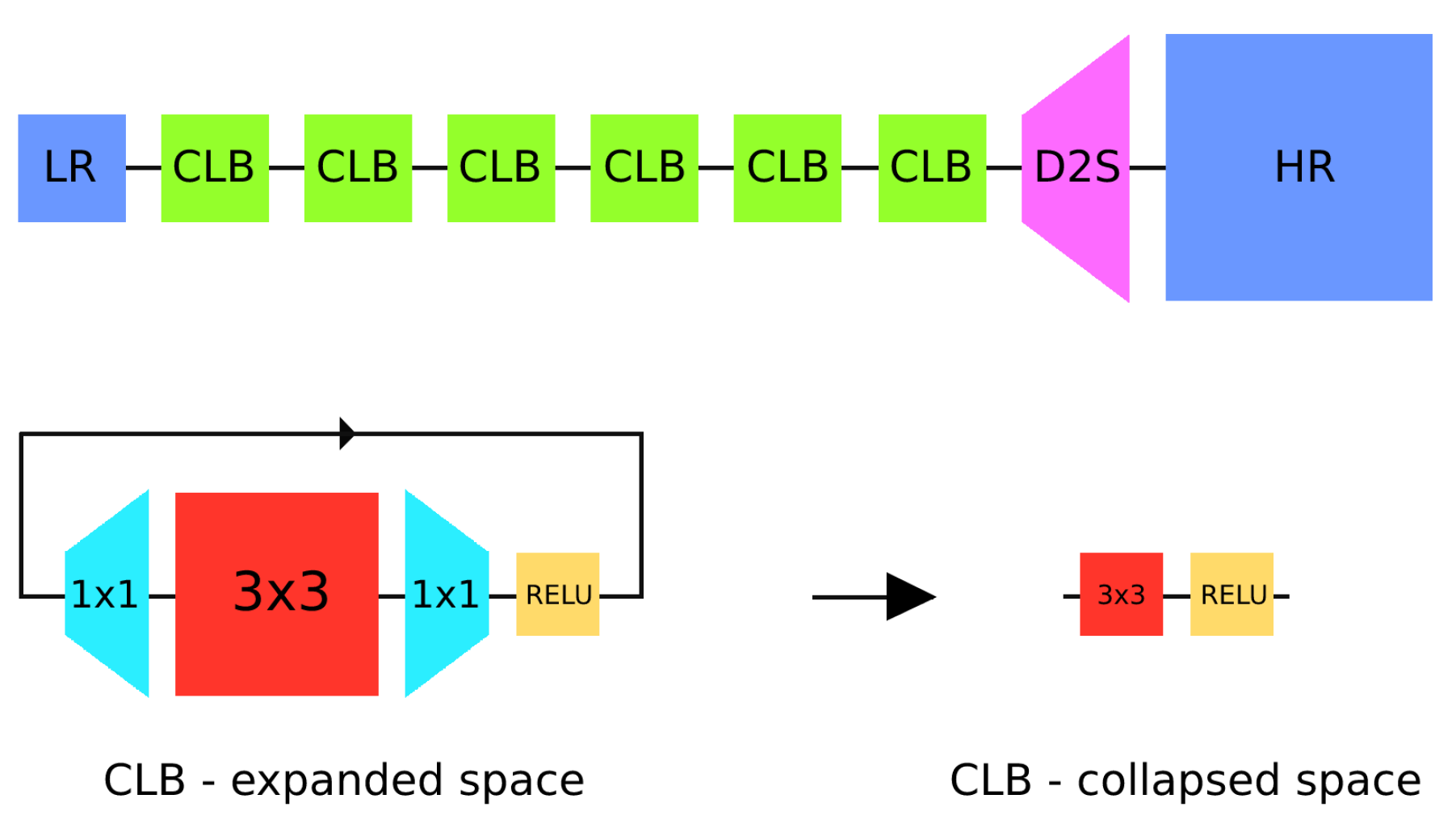}
\caption{\small{An overview of the model architecture and the structure of the CLB proposed by team TCLResearchEurope.}}
\label{fig:TCLResearchEurope}
\end{figure*}

The solution proposed by team TCLResearchEurope is based on two core ideas: Structural Reparametrization~\cite{zagoruyko2017diracnets,ding2021diverse,ding2021repvgg} and Channel Pruning~\cite{anwar2017structured,anwar2016compact}. To accelerate the training process, to limit the memory usage, and to make the quantization aware training in the expanded space possible, an online version of the Reparametrization was adopted~\cite{bhardwaj2022collapsible}. During the training phase, every convolution is represented as a Collapsible Linear Block (CLB) with skip connection (if convolution input channel number is equal to output channel number) with an initial expand ratio of 4. CLB employs similar ideas to the Inverted Residual Block~\cite{sandler2018mobilenetv2}, but the depth-wise convolution is replaced with the standard one in this block. The overall model architecture and the structure of the CLB block in both expanded and collapsed states are shown in Fig.~\ref{fig:TCLResearchEurope}. The initial number of input and output channels for intermediate convolution layers is set to 48. During the inference state, the model is collapsed into a plain convolutional structure with the depth-to-space layer preceded by the clip-by-value op. No residual connections were used in the model to keep its latency as small as possible, and ReLU was used as an activation function.

Channel Pruning was adopted to further improve the model efficiency. After each trained expanded convolution layer, a small two-layers Channel Ranker network was inserted. Such model is responsible for detecting the least important convolution channels within each layer. Later, during the distillation phase, such channels are, along with the Channel Ranker networks, removed and the model is further fine-tuned. Finally, the authors used the quantization aware training: since the proposed network has a simple feed-forward structure, post-training quantization leads to a significant accuracy drop. Quantizers were inserted right after each collapsed convolution block.

The model was trained on patches of size $128\times128$ pixels with a batch size of 16. The training process was done in three stages: first, network parameters were optimized for 1000 epochs using the Adam algorithm with a learning rate of 1e--4 decreased with the cosine scheduler. Next, the model was pruned and trained for another 1000 epochs. Finally, quantization aware training was performed for 100 epochs with a learning rate of 1e--7. During this stage, the authors used an additional loss function called the Self-supervised Quantization-aware Calibration Loss (SQCL)~\cite{wang2021fully}. This loss function is responsible for calibrating the values of parts of the network after quantization to have an approximate distribution as before quantization. SQCL allows to make the training process more stable and decrease the gap in PSNR before and after quantizing the model.

\subsection{ECNUSR}

\begin{figure*}[h!]
\centering
\includegraphics[width=0.5\linewidth]{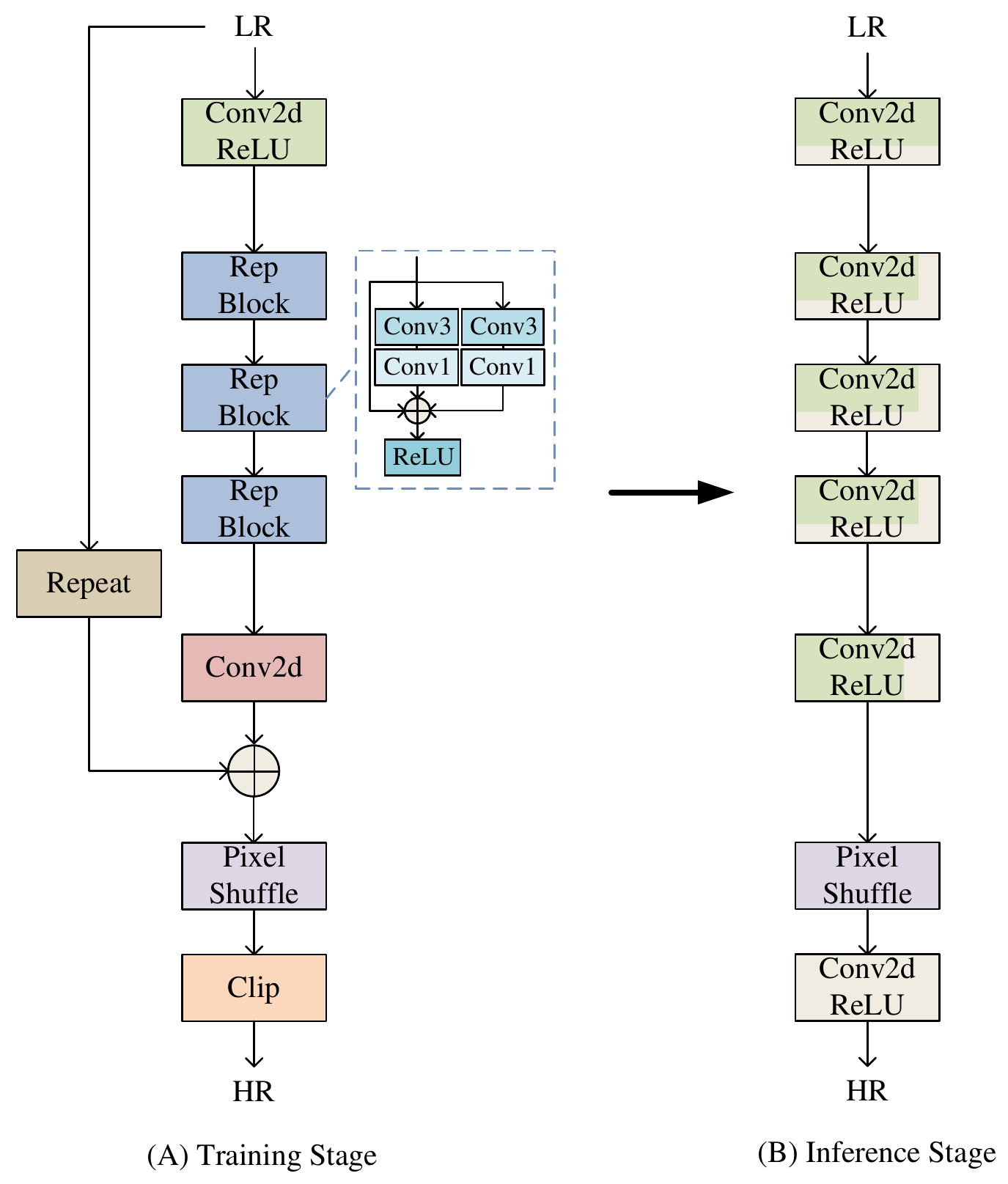}
\caption{\small{PureConvSR model proposed by team ECNUSR.}}
\label{fig:ECNUSR}
\end{figure*}

Team ECNUSR proposed a lightweight convolutional network with equivalent transformation (PureConvSR) shown in Fig.~\ref{fig:ECNUSR}. Equivalent transformation here refers to the replacement of time-consuming modules or operations such as \textit{clip}, \textit{add} and \textit{concatenate} with convolution modules. In the training stage, the network is composed of $N$ convolution layers, a global skip connection, a pixel shuffle module and a clip module. In the inference stage, the model is only composed of $N+1$ convolution layers and a pixel shuffle module. These two models are completely equivalent in logic due to the used equivalent transformation and reparameterization. The authors additionally used the EMA training strategy to stabilize the training process and improve the performance of the network. The model was trained using the MSE loss function on patches of size $64\times64$ pixels with a batch size of 16, network parameters were optimized using the Adam algorithm.

\subsection{LCVG}

\begin{figure*}[h!]
\centering
\includegraphics[width=0.9\linewidth]{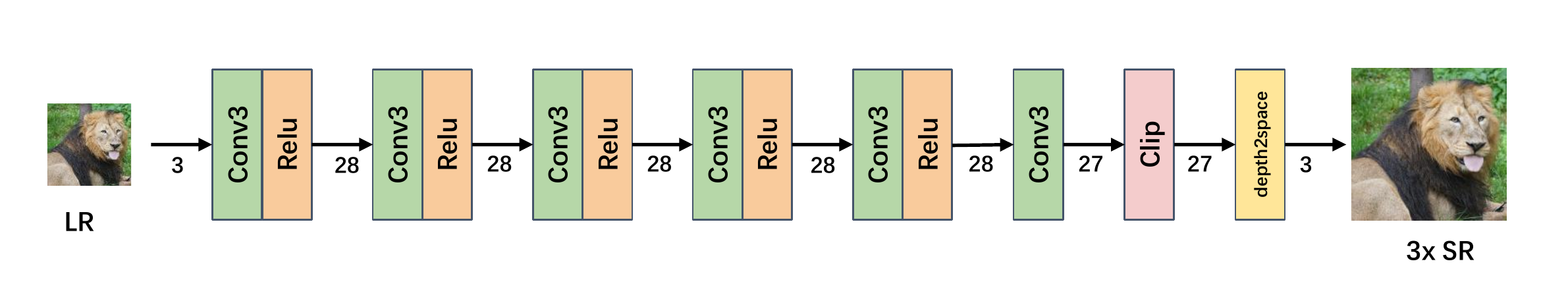}
\caption{\small{HOPN architecture developed by team LCVG.}}
\label{fig:LCVG}
\end{figure*}

Team LCVG proposed a simple ABPN~\cite{du2021anchor}-based network that consists of only $3\times3$ convolutions, clip node, and the depth to space operations (Fig.~\ref{fig:LCVG}). To improve its runtime, the authors removed the  multi-copy residual connection proposed in~\cite{du2021anchor} as the model without a residual connection demonstrated lower latency, although its PSNR score decrease was only marginal. The authors also exchanged the clip node and the depth-to-space op as this resulted in a better model runtime. Quantization aware training process was used when training the network. The authors found that QAT is sensitive to hyper-parameters and can easy fall into local optima if initializing INT8 model with pre-trained FP32 weights. Thus, to improve its PSNR score, quantized-aware training was performed from scratch. The model was trained to minimize the mean absolute loss (MAE) on patches of size $64\times64$ pixels with a batch size of 16. Network parameters were optimized for 1000K iterations using the Adam algorithm with a learning rate of 6e--4 decreased to 1e--8 with the cosine annealing strategy. Horizontal and vertical flips were used for data augmentation.

\subsection{BOE-IOT-AIBD}

\begin{figure*}[h!]
\centering
\includegraphics[width=0.9\linewidth]{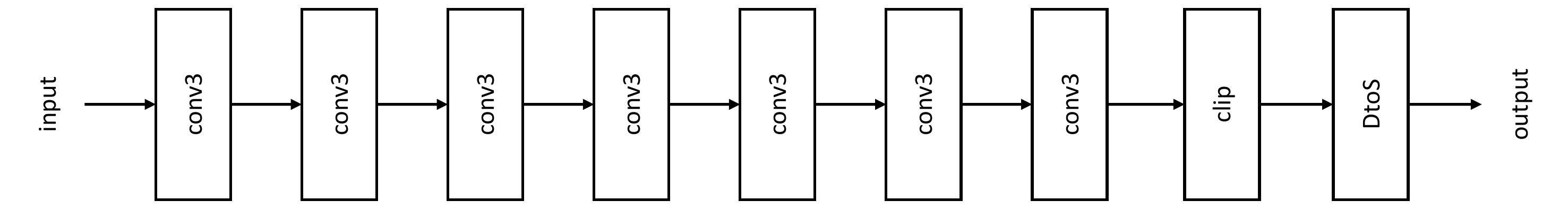}
\caption{\small{Model architecture proposed by team BOE.}}
\label{fig:BOE}
\end{figure*}

Team BOE designed a small CNN model that takes a low-resolution input image and passes it to a $3\times3$ convolutional layer performing feature extraction, followed by six $3\times3$ convolutions, one clip node and one depth-to-space layer. The authors found that placing the clip node before the depth-to-space layer is much faster than placing it after the depth-to-space layer. The model was trained to minimize the L1 loss on patches of size $128\times128$ pixels with a batch size of 8. Network parameters were optimized for 1000 epochs using the Adam algorithm with a learning rate of 1e--3 decreased by 0.6 every 100 epochs. Quantized-aware training was then applied to improve the accuracy of the resulting INT8 model. During this stage, the model was trained for another 220 epochs with the initial learning rate of 1e--5 decreased by 0.6 every 50 epochs. Random flips and rotations were used for data augmentation.

\subsection{NJUST}

\begin{figure*}[h!]
\centering
\includegraphics[width=0.8\linewidth]{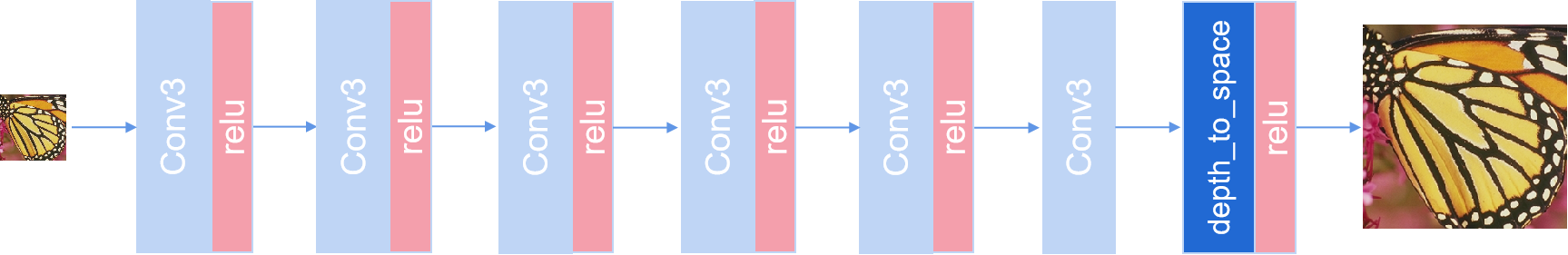}
\caption{\small{CNN architecture proposed by team NJUST.}}
\label{fig:NJUST}
\end{figure*}

The model architecture developed by team NJUST is very similar to the previous two solutions: the network contains five $3\times3$ convolutional layers followed by ReLU activations, and one depth-to-space layer (Fig.~\ref{fig:NJUST}). No skip connections were used to improve the model efficiency. The network was trained using the L1 loss function on patches of size $64\times64$ pixels with a mini-batch size of 16. Model parameters were optimized for 600K iterations with a learning rate of 1e--3 halved every 150K iterations. Quantized-aware training was applied to improve the accuracy of the resulting INT8 model.

\subsection{Antins\_cv}

\begin{figure*}[h!]
\centering
\includegraphics[width=0.9\linewidth]{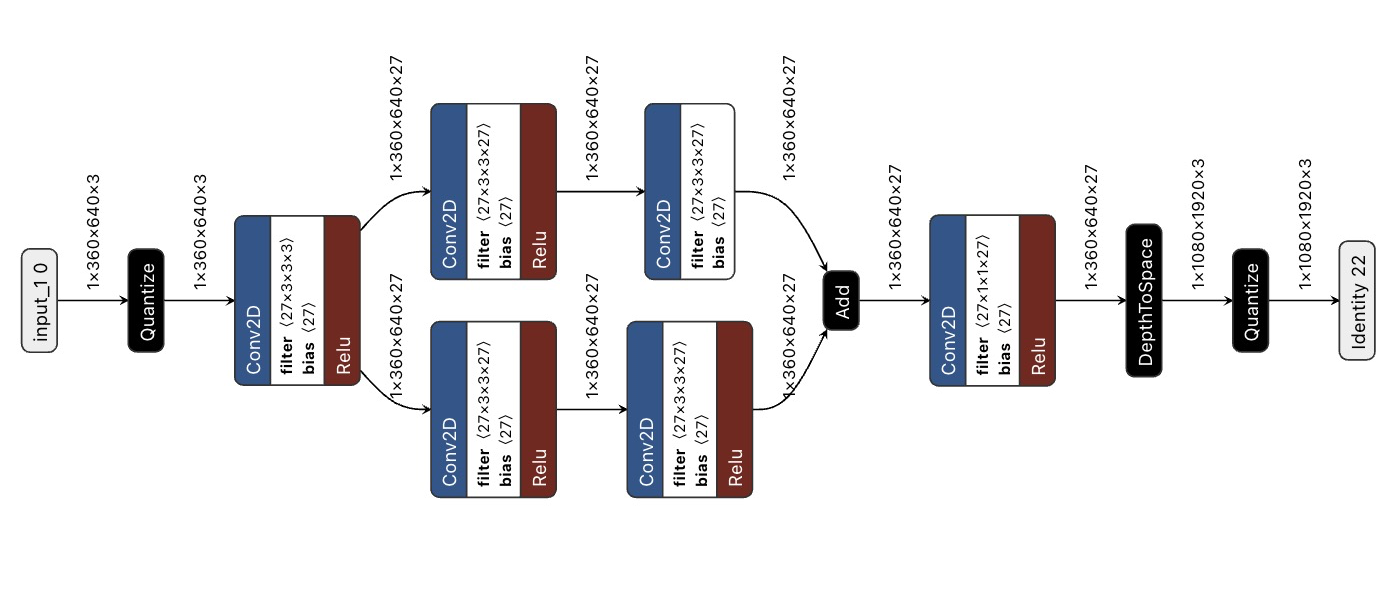}
\caption{\small{Antins\_cv.}}
\label{fig:Antins}
\end{figure*}

The architecture designed by team Antins\_cv is demonstrated in Fig.~\ref{fig:Antins}. The model consists of six convolutional layers: after the first convolutional layer, the resulting features are then processed by two network branches. Each branch has two convolutional layers, and the second layer of right right branch is not followed by any activation function. The outputs of these branches are merged pixel-wise using the add operation as it turned to be faster than the concat op. The authors used a pixel shuffle layer to produce the final output image.

The model was first trained to minimize the PSNR loss function on patches of size $128\times128$ pixels with a batch size of 32. Network parameters were optimized for 600 epochs using the Adam optimizer with a learning rate of 1e--3 halved every 120 epochs. Then, the authors applied quantization-aware training: MSE loss was used as the target metric, the batch size was set to 8, and the model was fine-tuned for 220 epochs with the initial learning rate of 1e--4 decreased by half every 50 epochs. The authors additionally proposed a new MovingAvgWithFixedQuantize op to deal with incorrect normalization of the model output caused by INT8 quantization.

\subsection{GenMedia Group}

\begin{figure*}[h!]
\centering
\includegraphics[width=0.9\linewidth]{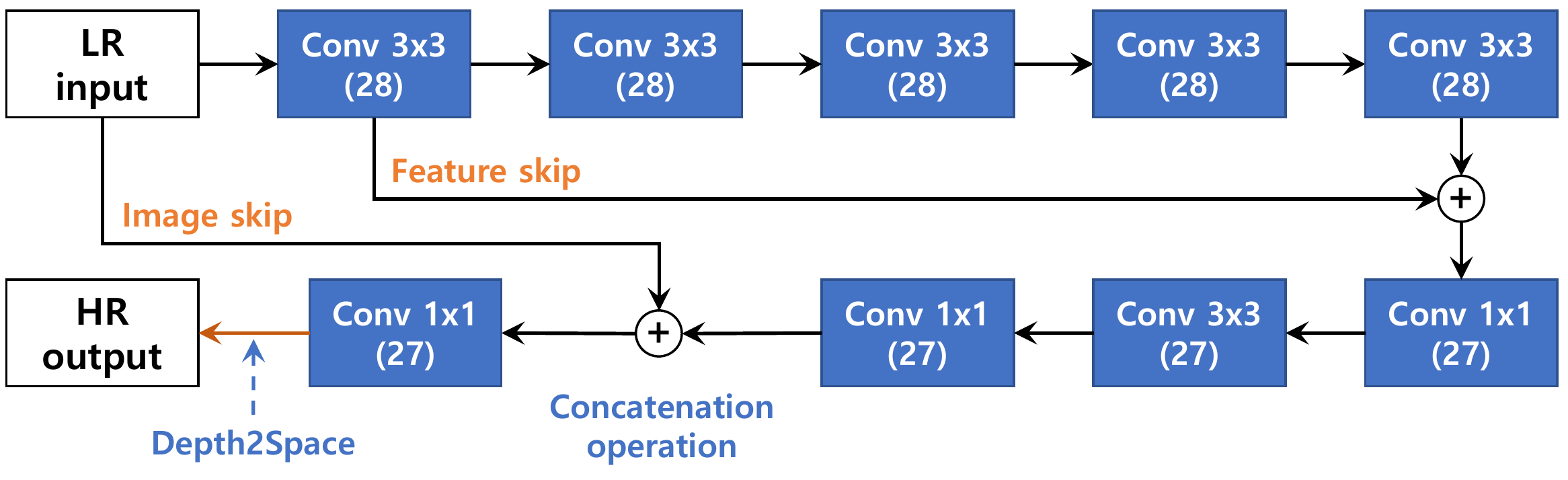}
\caption{\small{A modified anchor-based plain network proposed by team GenMedia Group.}}
\label{fig:GenMedia}
\end{figure*}

The model architecture proposed by team GenMedia Group is also inspired by the last year's top solution~\cite{du2021anchor}. The authors added one extra skip connection to the mentioned anchor-based plain net (ABPN) model, and used concatenation followed by a $1\times$1 convolution instead of a pixel-wise summation at the end of the network (Fig.~\ref{fig:GenMedia}). To avoid potential context switching between CPU and GPU, the order of the clipping function and the depth-to-space operation was rearranged. The authors used the mean absolute error (MAE) as the target objective function. The model was trained on patches of size $96\times96$ pixels with a batch size of 32. Network parameters were optimized for 1000 epochs using the Adam algorithm with a learning rate of 1e--3 decreased by half every 200 epochs. For FP32 training, the authors used SiLU activation function that was later replaced with ReLU when fine-tuning the model with quantization-aware training. Random horizontal and vertical image flips were used for data augmentation.

\subsection{Vccip}

\begin{figure*}[h!]
\centering
\includegraphics[width=1.0\linewidth]{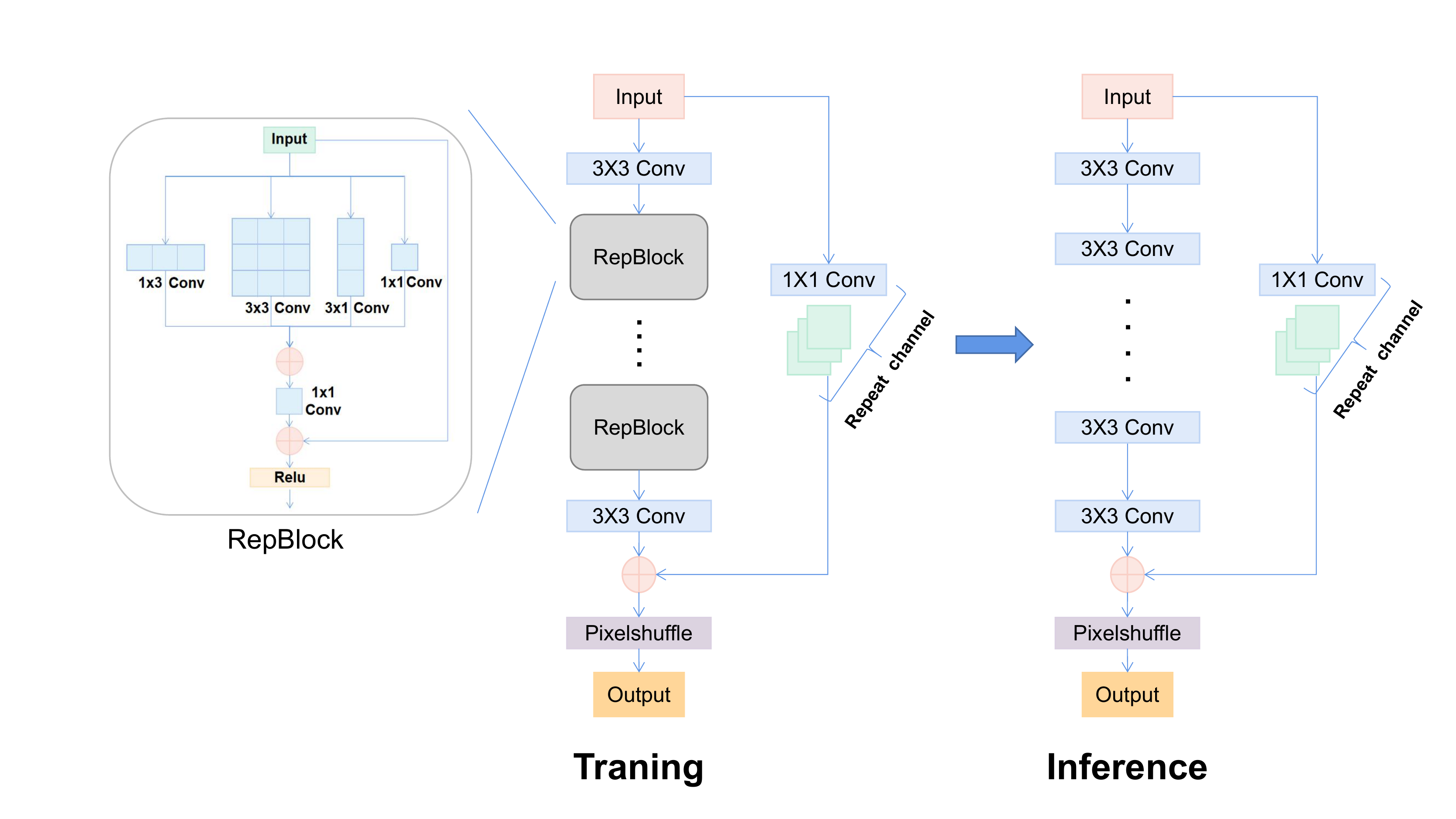}
\caption{\small{The architecture of the model and the structure of the Re-Parameterizable blocks (RepBlocks) proposed by team Vccip.}}
\label{fig:Vccip}
\end{figure*}

Team Vccip derived their solution from the ABPN~\cite{du2021anchor} architecture. The proposed model consists of two $3\times3$ convolution layers, six Re-Parameterizable blocks (RepBlocks), and one pixel shuffle layer (Fig.~\ref{fig:Vccip}). The RepBlock can enhance the representational capacity of a single convolutional layer by combining diverse branches of different scales, including sequences of convolutions, multi-scale convolutions and residual structures. In the inference stage, the RepBlock is re-parameterized into a single convolutional layer for a better latency. As in~\cite{du2021anchor}, only a residual part of the SR image is learned. The model was first trained to minimize the Charbonnier and Vgg-based perceptual loss functions on patches of size $64\times64$ pixels with a batch size of 16. Network parameters were optimized for 600K iterations using the Adam optimizer with the cyclic learning rate policy. Quantization-aware training was applied to improve the accuracy of the resulting INT8 model.

\subsection{MegSR}

\begin{figure*}[h!]
\centering
\includegraphics[width=0.9\linewidth]{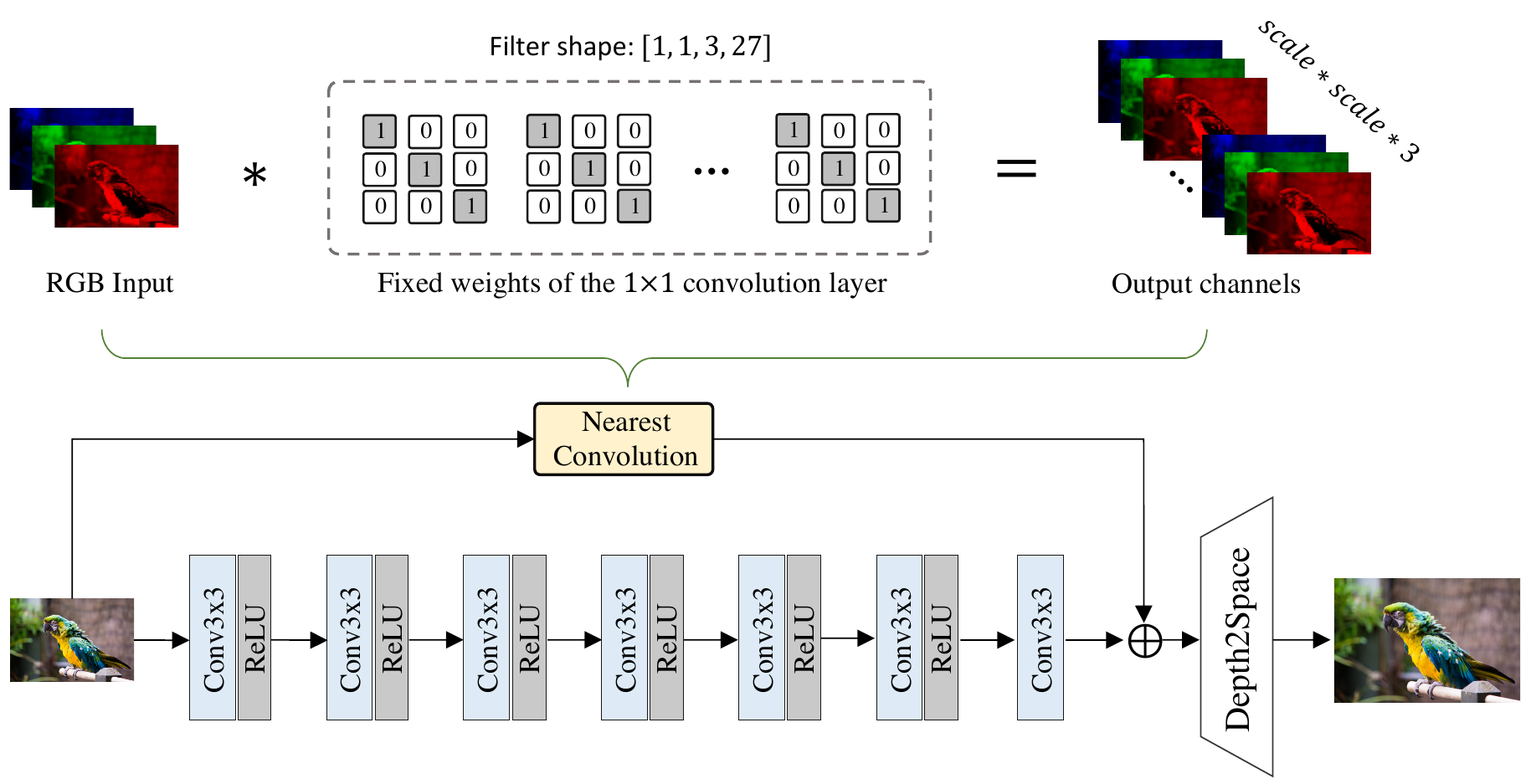}
\caption{\small{NCNet model proposed by team MegSR.}}
\label{fig:MegSR}
\end{figure*}

The architecture of the Nearest Convolution Network (NCNet)~\cite{luo2022fast} proposed by team MegSR is demonstrated in Fig.~\ref{fig:MegSR}. This model constrains a Nearest Convolution module, which is technically a special $1\times1$ convolution layer with a stride of 1. To achieve the nearest interpolation, the weights of this module are freezed and manually filled by $s^2$ groups of $3\times3$ identity matrix (where $s$ is the upscaling factor), and each group would produce an RGB image, performing a copy operation to repeat the input image. When followed by the depth-to-space operation, this module would reconstruct exactly the same image as in case of the nearest interpolation method, though much faster when running on mobile GPUs or NPUs. 

The entire model consists of seven $3\times3$ convolution layers with ReLU activations, the number of channels is set to $32$. The model was trained to minimize the L1 loss function on patches of size $64\times64$ pixels with a batch size of 64. Network parameters were optimized for 500K iterations using the Adam optimizer with the initial learning rate of 1e--3 decreased by half every 200K iterations. Inspired by~\cite{li2022ntire}, the authors additionally fine-tuned the model on larger patch size of $128\times128$ for another 200K iterations.

\subsection{DoubleZ}

\begin{figure*}[h!]
\centering
\includegraphics[width=0.7\linewidth]{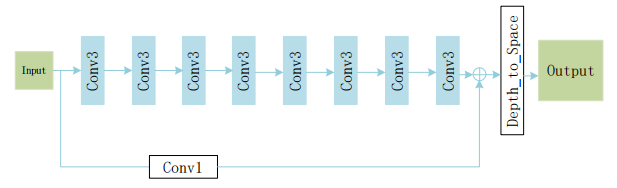}
\caption{\small{ABPN-based network proposed by team DoubleZ.}}
\label{fig:DoubleZ}
\end{figure*}

The architecture proposed by team DoubleZ is very similar to the ABPN~\cite{du2021anchor} network: the major difference consists in the residual module, where one convolution layer is used instead of 9 times input image stacking. The authors used ReLU activation to clip the outputs of the model instead of the clip\_by\_value op as it turned to be faster in the experiments. The model was trained in two stages: first, it was minimizing the L1 loss function on $96\times96$ pixel patches using the Adam optimizer with a learning rate of 1e--3 decreased to 1e--5 through the training. Then, the model was fine-tuned with the L2 loss. Random flips and rotations were used for data augmentation, quantization-aware training was applied to improve the accuracy of the resulting INT8 model.

\subsection{Jeremy Kwon}

\begin{figure*}[h!]
\centering
\includegraphics[width=0.7\linewidth]{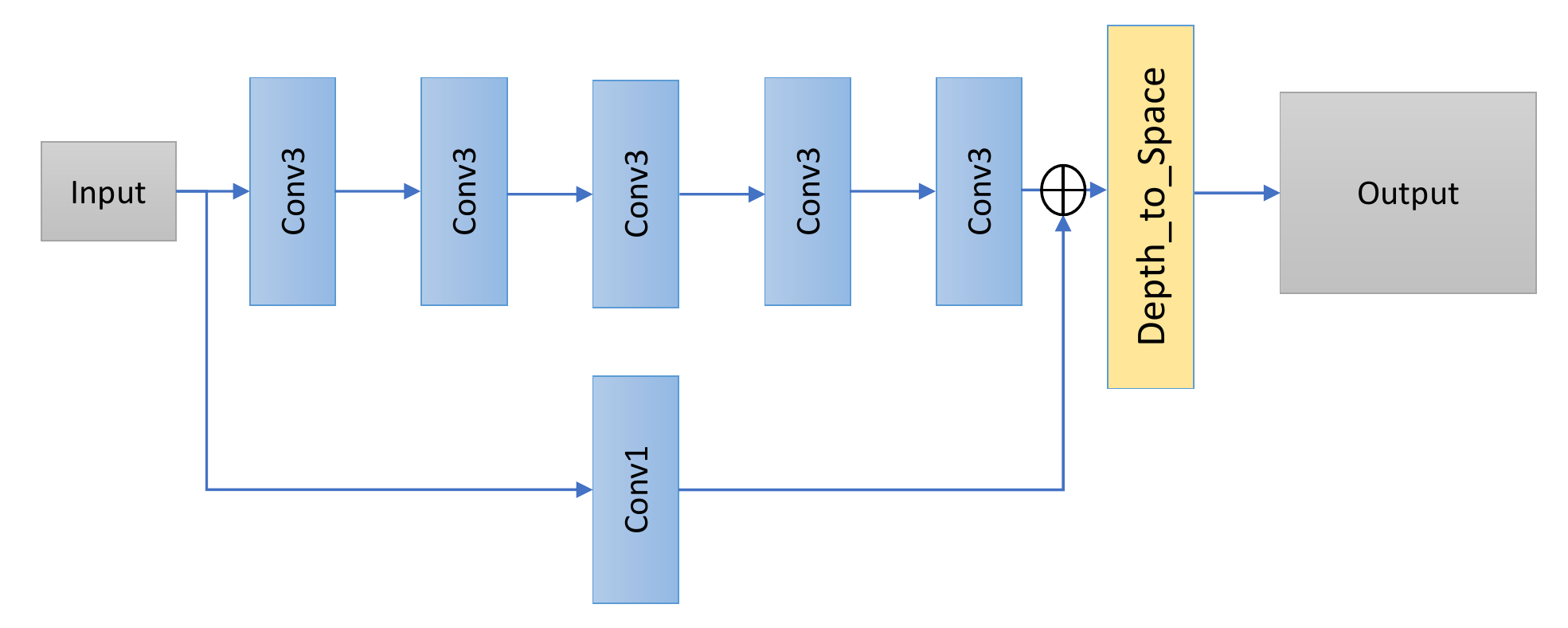}
\caption{\small{ABPN-based network proposed by team Jeremy Kwon.}}
\label{fig:Jeremy}
\end{figure*}

Jeremy Kwon used almost the same ABPN-based model design as the previous team (Fig.~\ref{fig:Jeremy}), the main difference between these solutions is the number of convolutional layers. The model was trained to minimize the L1 and MAE losses (the latter one was computed on images resized with the space-to-depth op) on patches of size $96\times96$ pixels with a batch size of 16. Network parameters were optimized for 1000 epochs using the Adam optimizer with the initial learning rate of 1e--3 decreased by half every 200 epochs. Quantization-aware training was applied to improve the accuracy of the resulting INT8 model.

\subsection{Lab216}

\begin{figure*}[h!]
\centering
\includegraphics[width=0.9\linewidth]{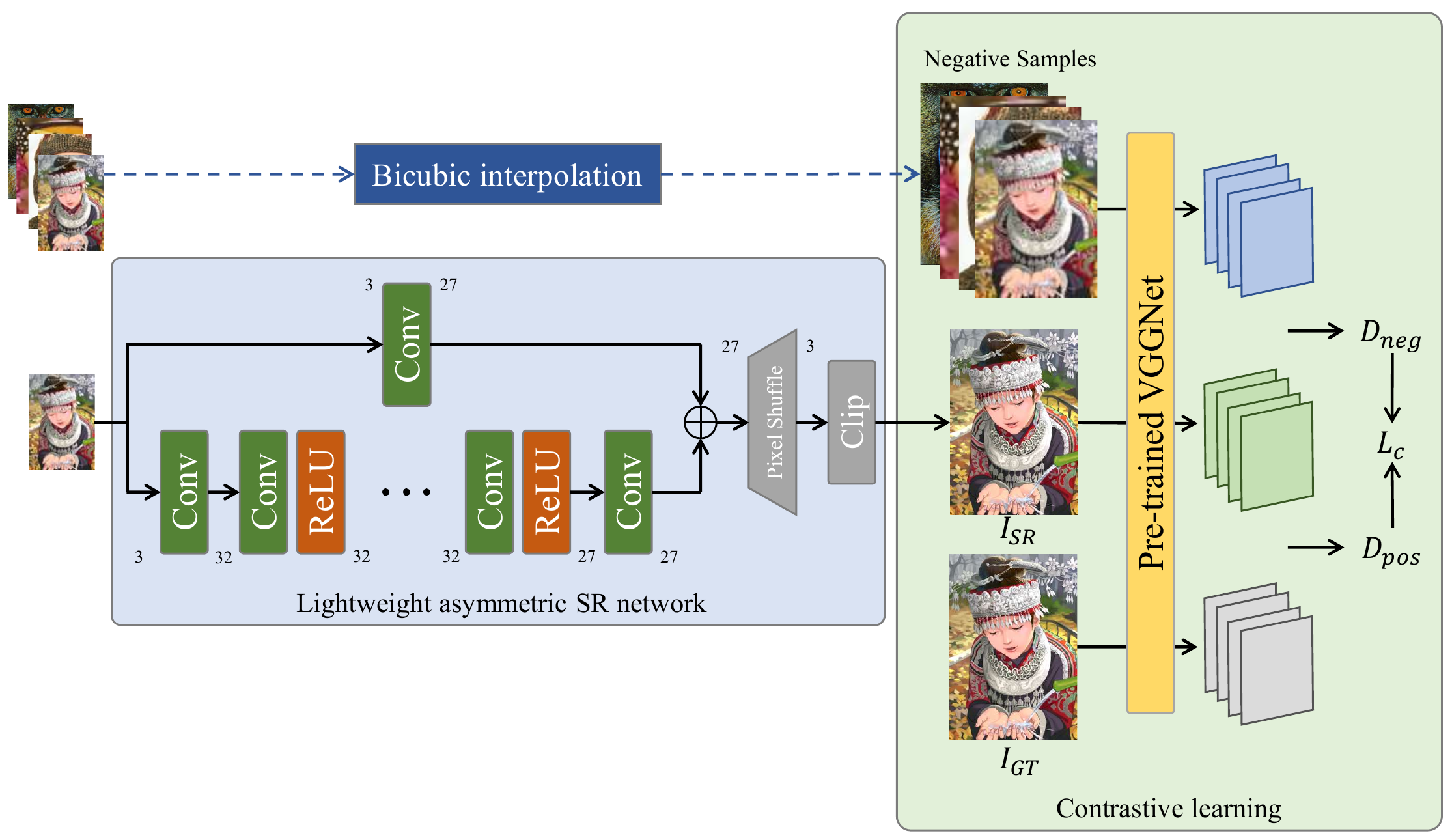}
\caption{\small{Model architecture trained with the contrastive quantization-aware method proposed by team Lab216.}}
\label{fig:Lab216}
\end{figure*}

Team Lab216 proposed a lightweight asymmetric network and applied the contrastive quantization-aware training to improve its performance after quantization (Fig.~\ref{fig:Lab216}). The model consists of two branches: the first residual branch has only one $3\times3$ convolution layer to extract basic features required for image up-sampling, while the second branch contains seven convolution layers with ReLU activiations to learn a more detailed information. The features extracted by these branches are summed up and then up-sampled with a pixel shuffle op.

The model was trained in two stages: first, the network was minimizing the L1 loss on patches of size $64\times64$ pixels with a batch size of 16. Network parameters were optimized for 1000 epochs using the Adam optimizer with the initial learning rate of 1e--3 decreased by half every 200 epochs.

Next, the authors used the contrastive loss~\cite{wang2021towards,kong2022residual} to boost the performance of quantization-aware training. It works by keeping the anchor close to positive samples while pushing away from negative samples in the considered latent space. In this case, the ground truths images are settled as positive samples, while negative sets are generated with bicubic interpolation. The total loss is defined as:
\begin{equation*}
\mathcal{L}_{total} = \mathcal{L}_1 + \lambda \times \mathcal{L}_c,
\end{equation*}
\begin{equation*}
\mathcal{L}_c = \frac{d(\phi(I_{SR}),\phi(I_{GT}))}{\sum_{k=1}^Kd(\phi(I_{SR}),\phi(I_{Neg}^k))},
\end{equation*}
where $K=16$ is the number of negative samples, $\phi()$ are the intermediate features generated by a pre-trained VGG-19 network. At this stage, the model was trained for 220 epochs with a learning rate initialized at 1e--4 and decreases by half every 50 epochs.

\subsection{TOVB}

\begin{figure*}[h!]
\centering
\includegraphics[width=0.9\linewidth]{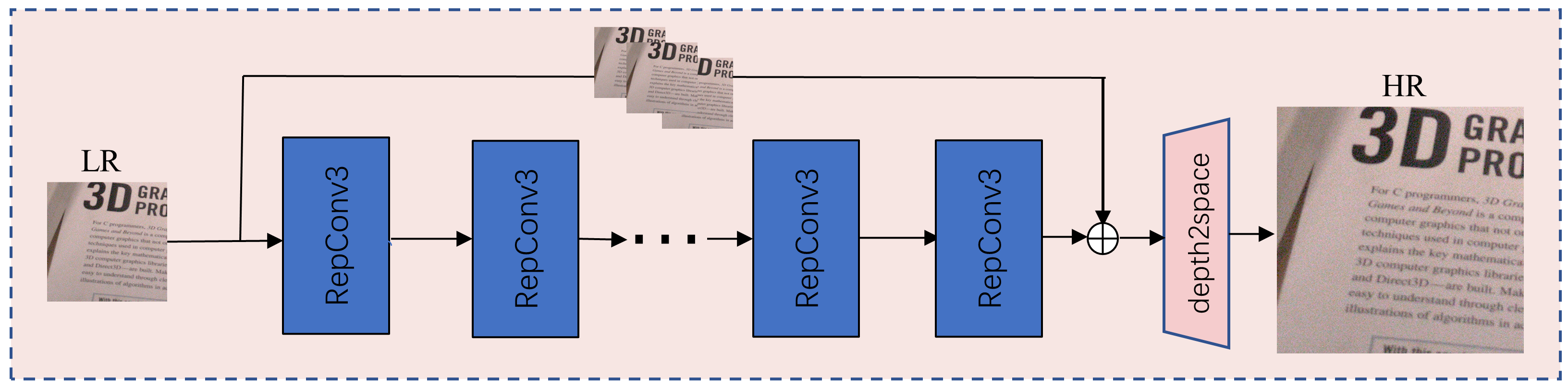}
\includegraphics[width=0.7\linewidth]{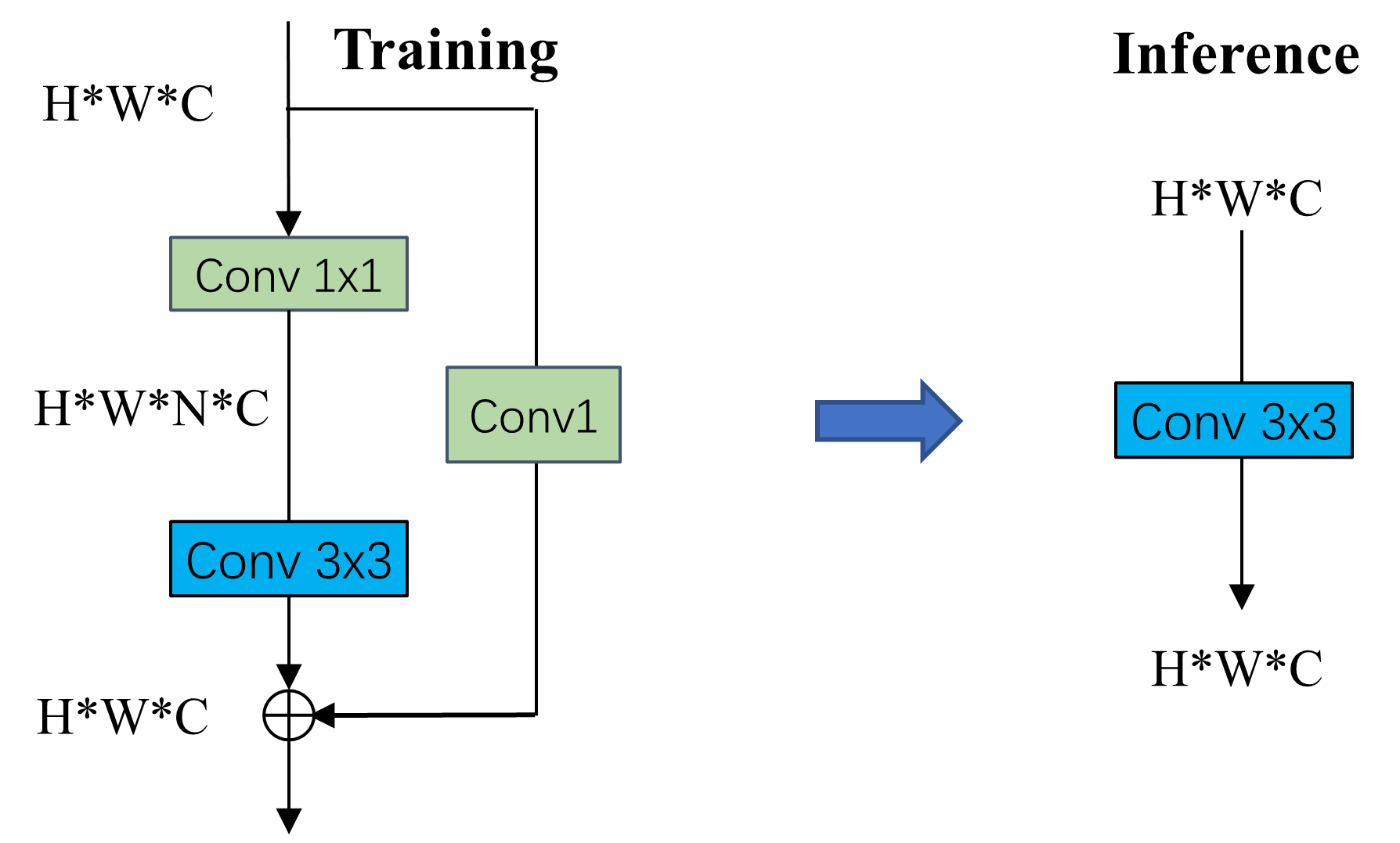}
\caption{\small{The overall model architecture and the structure of the RepConv block proposed by team TOVB.}}
\label{fig:TOVB}
\end{figure*}

Team TOVB proposed an ABPN-based model architecture, where $3\times3$ convolutions were replaced with a re-parameterized convolution module (Fig.~\ref{fig:TOVB}). In total, the network consists of seven RepConv modules. The model was trained in two stages: first, L1 loss was used as the target metric, and model parameters were optimized for 1500 epochs using the Adam algorithm with a batch size of 64 and a learning rate of 2e--4 on patches of size $160\times160$. In the second stage, each RepConv module was fused into a single convolution layer, and the model was fine-tuned for another 30 epochs on patches of size $64\times64$ with a batch size of 16 and a learning rate of 1e--6.

\subsection{Samsung Research}

Team Samsung Research used the ABPN~\cite{du2021anchor} network design. First, the authors trained the model on the ImageNet-2012 data. Then, the network was fine-tuned on the DIV2K and Flickr2K data. Finally, the model was further fine-tuned using the quantization-aware training. The difference of this solution compared from the original ABPN paper is that a smaller learning rate of 1e--5 was used during the quantization-aware training step.

\subsection{Rtsisr2022}

\begin{figure*}[h!]
\centering
\includegraphics[width=1.0\linewidth]{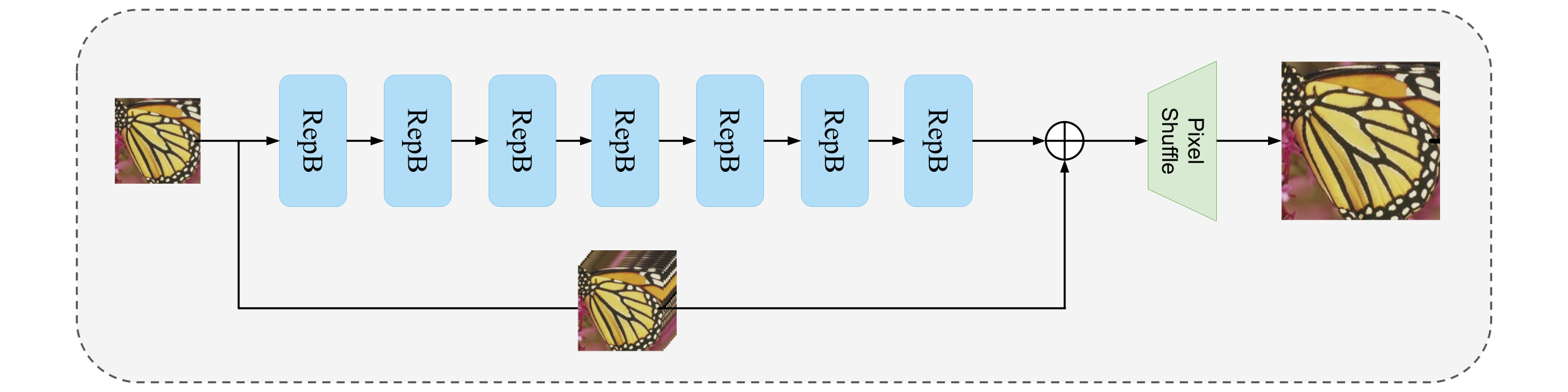}
\includegraphics[width=0.6\linewidth]{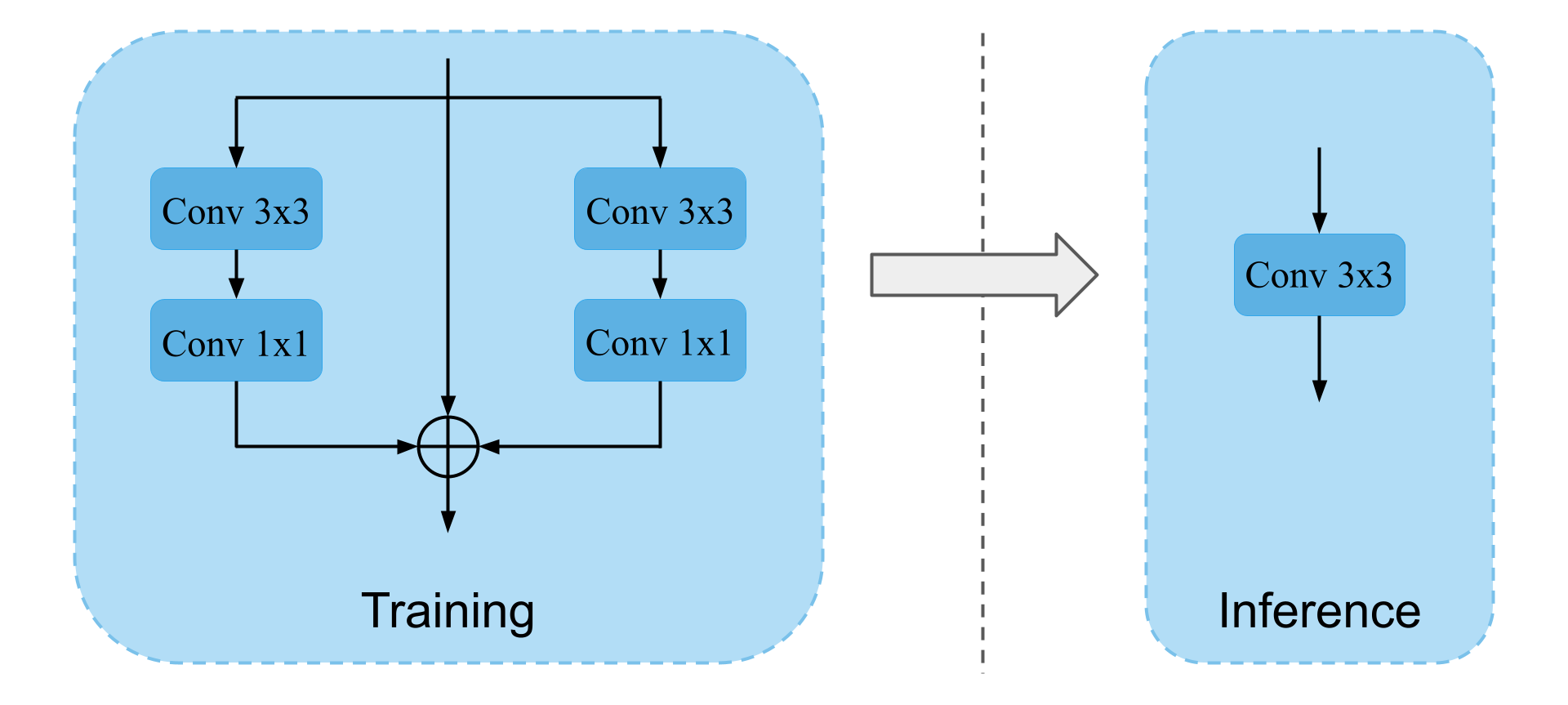}
\caption{\small{The overall model architecture and the structure of the RepB block proposed by team Rtsisr2022.}}
\label{fig:Rtsisr2022}
\end{figure*}

Team Rtsisr2022 adopted the ABPN architecture, but changed the number of channels from 28 to 32 as the model is in general more efficiently executed on NPUs when the number of channels is a multiple of 16. Furthermore, to improve the accuracy results while keeping the same runtime, the authors replaced every $3\times3$ convolutional layer with a re-parameterize convolution block (RepB) illustrated in Fig.~\ref{fig:Rtsisr2022}.

The model was trained in three stages: first, L1 loss was used as the target metric, and model parameters were optimized for 2000 epochs using the Adam algorithm with a batch size of 16 and a learning rate of 6e--4 decayed by half every 200 epochs. In the second stage, each RepB module was fused into a single convolution layer, and the model was fine-tuned for another 1000 with a learning rate of 8e--5 decreased by half every 200 epochs. Finally, the authors applied quantization-aware training to improve the accuracy of the resulting INT8 model. At this stage, the model was fine-tuned for 220 epochs with a learning rate of 8e--6 decreased by half every 50 epochs. Patches of size $192\times192$ augmented with random flips and rotations were used at all stages.

\subsection{Aselsan Research}

\begin{figure*}[h!]
\centering
\includegraphics[width=0.9\linewidth]{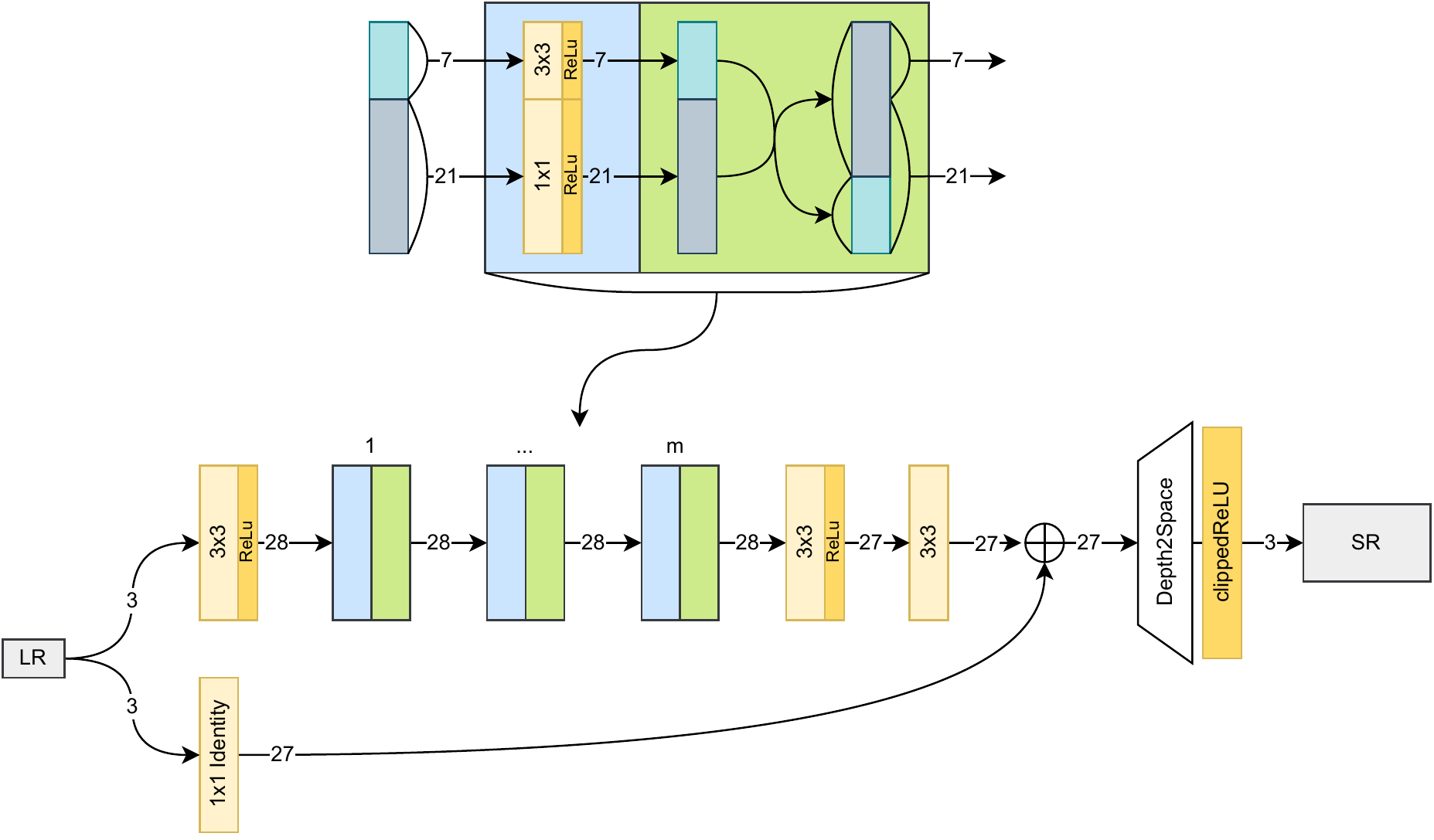}
\caption{\small{XCAT model architecture proposed by team Aselsan Research.}}
\label{fig:Aselsan}
\end{figure*}

Team Aselsan Research proposed the XCAT model~\cite{ayazoglu2022xcat} demonstrated in Fig.~\ref{fig:Aselsan}, its distinctive features and their impact on the performance are as follows:

\begin{itemize}
\item \textBF{Group convolutions with unequal filter groups and dynamic kernels.}
Group convolutions~\cite{krizhevsky2012imagenet} include multiple kernels per layer to extract and learn more varying features, compared to a single convolutional layer. Inspired by XLSR~\cite{ayazoglu2021extremely}'s GConv blocks, XCAT also has repeated group convolution blocks to replace convolutional layers. However, convolutional layers inside the group convolution blocks in XCAT have different layer dimensions and dynamic kernel sizes. This allows to pass the same source information between differently specified convolutional layers and allows for a better feature extraction. 
\item \textBF{Cross concatenation.} Instead of using $1\times1$ depthwise convolutions for increasing the spatial receptive field of the network as done in XLSR's GConv blocks, the output tensor of each group convolution block is circularly shifted.

\item \textBF{Intensity-augmented training.} To minimize PSNR difference between the original FP32 and quantized INT8 models, intensity values of the training images are scaled with a randomly chosen constant (1, 0.7, 0.5).

\item \textBF{Nearest neighborhood up sampling with fixed kernel convolutions.} The authors observed that providing a low resolution input image to the depth-to-space op with accompanying feature tensors increases the PSNR score as opposed to only providing the extracted feature tensors. With this motivation and the inspiration from~\cite{du2021anchor}, XCAT adds a repeated input image tensor (where each channel repeated 9 times) to feature tensors, and provides it to the depth-to-space op. For a better performance, a convolutional layer with 3 input channels and 27 output channels is used instead of the tensor copy op, with a $1\times1$ untrainable kernel set to serve the same purpose.

\item \textBF{Clipped ReLU.} Using clipped ReLU at the end of the model allows to deal with incorrect output normalization when performing model quantization.
\end{itemize}

The model was trained twice. First, the authors used the Charbonnier loss function and optimized model parameters using the Adam algorithm with the initial learning rate of 1e--3 and the warm-up scheduler. During the second stage, the MSE loss function was used, and the initial learning rate was set to 1e--4.

\subsection{Klab\_SR}

\begin{figure*}[h!]
\centering
\includegraphics[width=1.0\linewidth]{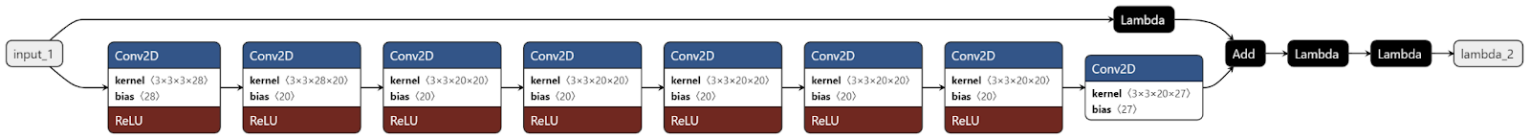}
\caption{\small{ABPN-based network proposed by team Klab\_SR.}}
\label{fig:Klab}
\end{figure*}

Team Klab\_SR proposed a modified ABPN-based network: the channel size was decreased from 28 to 20, while two additional convolution layers were added for a better performance (Fig.~\ref{fig:Klab}). Model parameters were optimized using the NadaBelief optimizer (a combination of the Adabelief~\cite{zhuang2020adabelief} and Nadam~\cite{dozat2016incorporating}), quantization-aware training was applied to improve the accuracy of the resulting INT8 model.

\subsection{TCL Research HK}

\begin{figure*}[h!]
\centering
\includegraphics[width=0.9\linewidth]{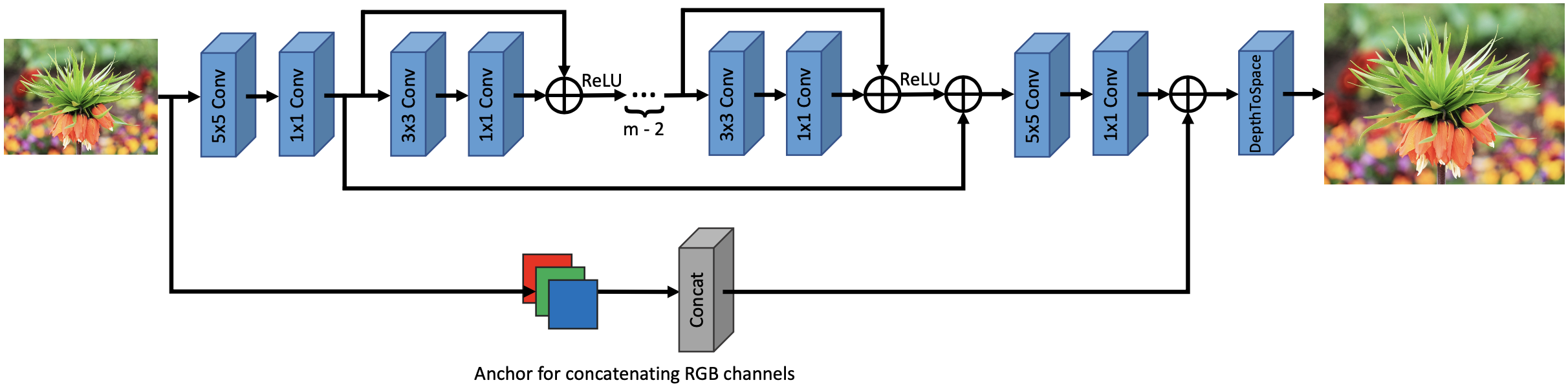}
\includegraphics[width=0.7\linewidth]{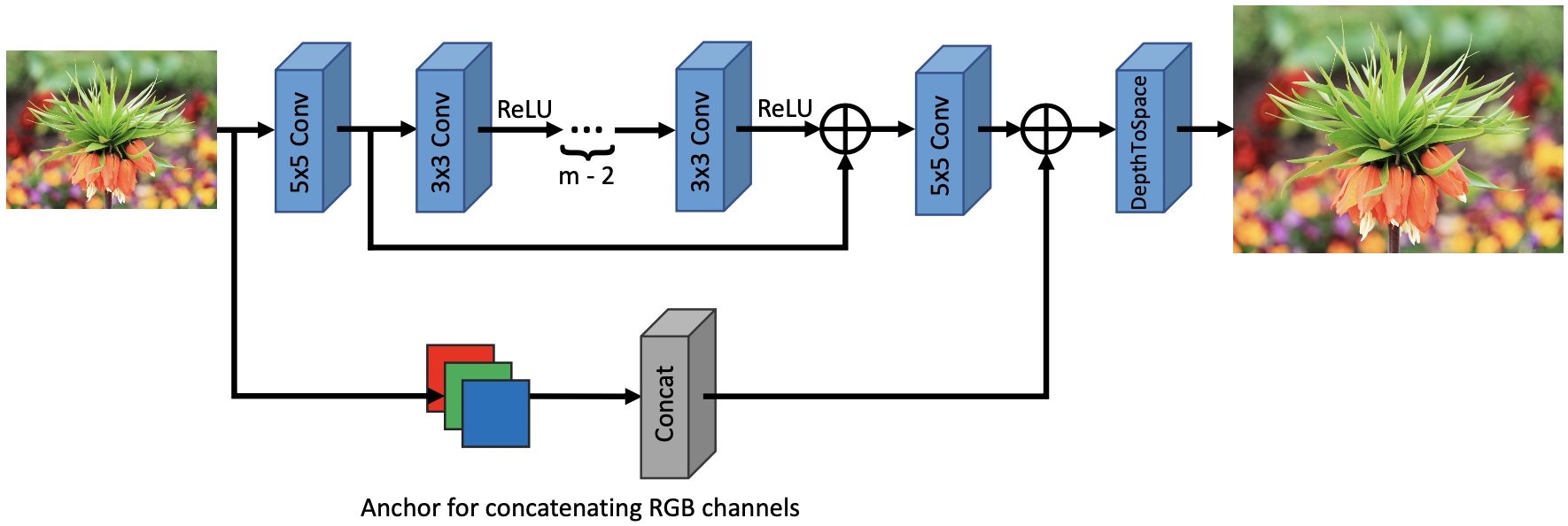}
\caption{\small{ACSR model architecture proposed by TCL Research HK during the training (top) and inference (bottom) stages.}}
\label{fig:TCL}
\end{figure*}

Team TCL Research HK proposed an anchor-collapsed super resolution (ACSR) network. The model contains the anchor, the feature extraction and the reconstruction modules as shown in Fig.~\ref{fig:TCL}. The anchor is a concatenation skip connection layer that stacks the images 9 times in order to learn the image residuals~\cite{du2021anchor}. The feature extraction module consists of several residual blocks with $3\times3$ followed by $1\times1$ convolutional layers in addition to a ReLU activation layer. The reconstruction module is a $5\times5$ followed by $1\times1$ convolutional layer with the addition of a pixel shuffle layer. After training, all blocks are collapsed into single convolutional layers~\cite{bhardwaj2022collapsible} to obtain a faster model.

ACSR model was trained to minimize the L1 loss function on patches of size $200\times200$ pixels with a batch size of 4. Network parameters were optimized for 1000 epochs using the Adam optimizer with the initial learning rate of 1e--4. Random flips and rotations were used for data augmentation, quantization-aware training was applied to improve the accuracy of the resulting INT8 model.

\subsection{RepGSR}

\begin{figure*}[h!]
\centering
\includegraphics[width=0.9\linewidth]{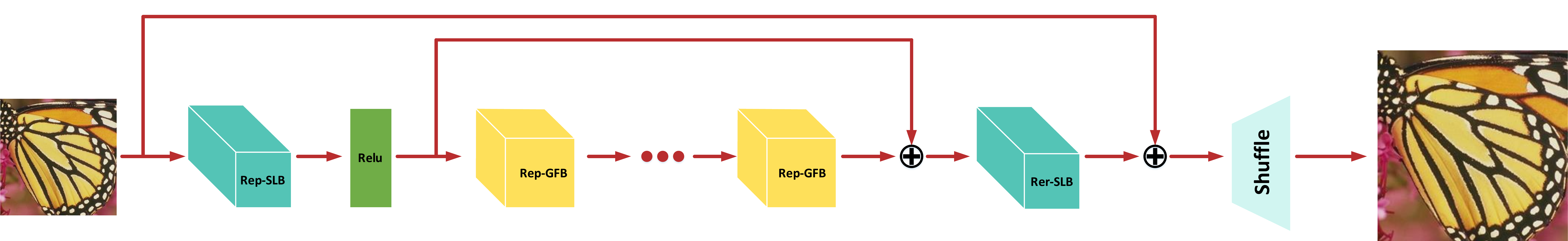}
\includegraphics[width=0.7\linewidth]{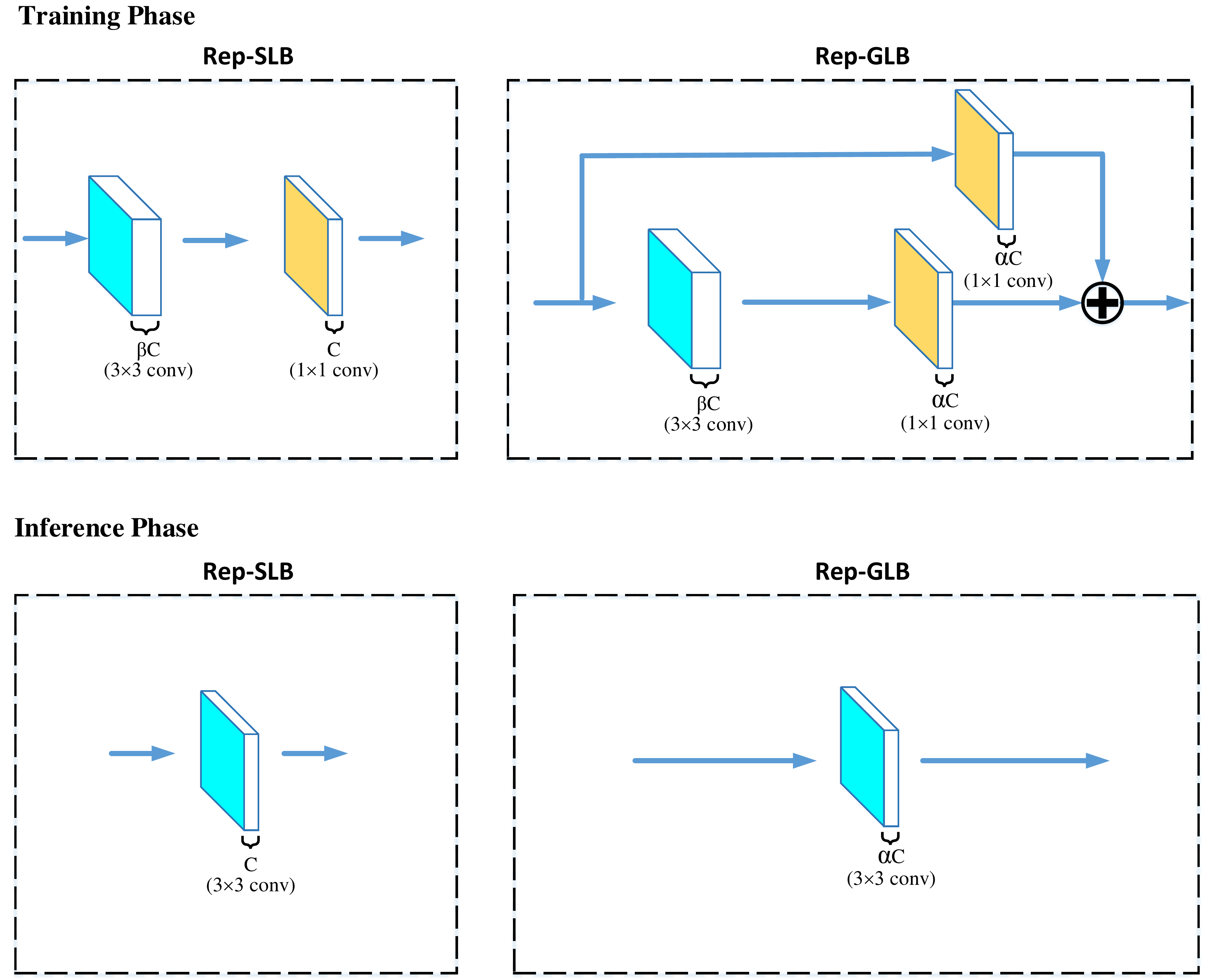}
\includegraphics[width=0.69\linewidth]{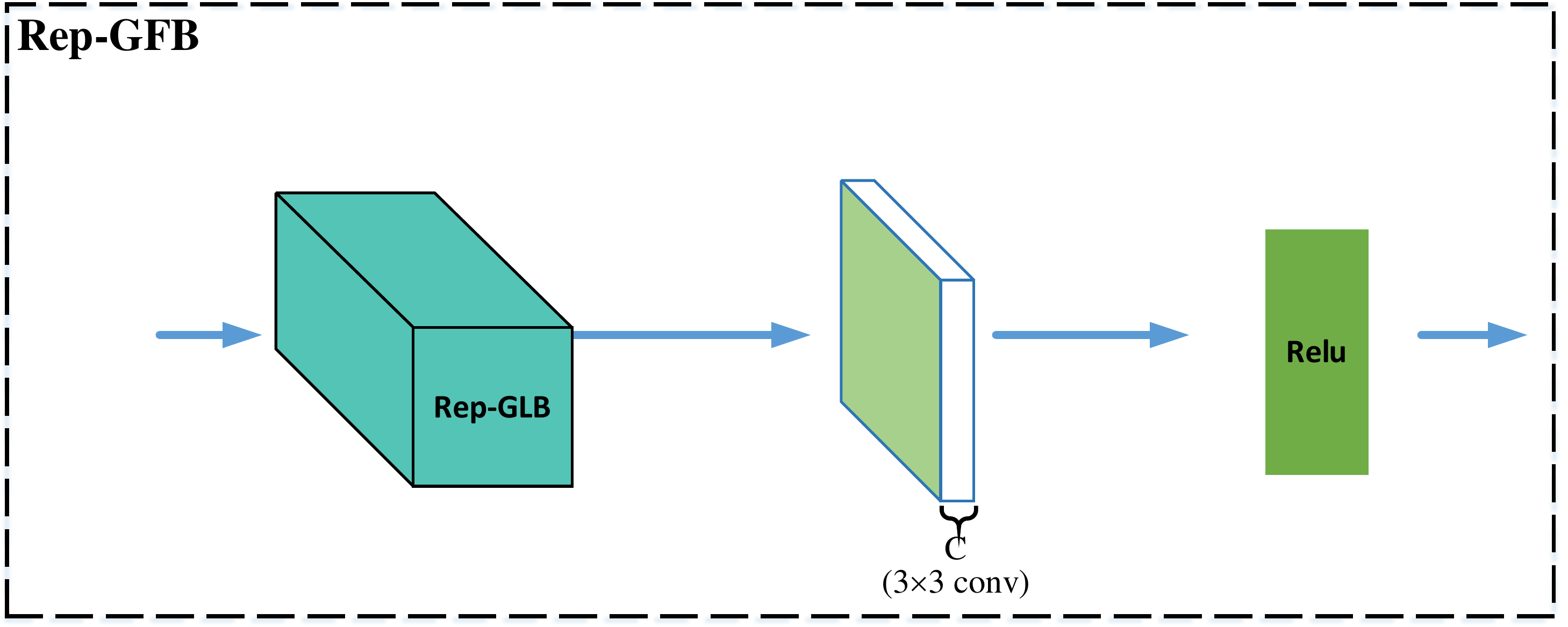}
\caption{\small{The overall model architecture (top), the structure of the Rep-LB (middle) and Rep-GFB (bottom) blocks proposed by team RepGSR.}}
\label{fig:RepGSR}
\end{figure*}

To obtain a lightweight super-resolution model, team RepGSR designed a network based on ghost features with re-parametarization. The model is comprised of two module types [Rep-Linear Blocks (Rep-LBs) and Rep-Ghost Features Blocks (Rep-GFBs)] and two block types [Rep-Simple Linear Blocks (Rep-SLBs) and Rep-Ghost Linear Blocks (Rep-GLBs)], which structure is demonstrated in Fig.~\ref{fig:RepGSR}. The input image is first passed to the rep-SLB block followed by ReLU activation that extracts shallow features. These features are then fed and processed by the following $m$ rep-GFB modules and one rep-SLB block. Finally, a pixel shuffle layer is used to get the final output SR image. During the inference, Rep-SLB and Rep-GLBs blocks are compressed using layer re-parameterization.

The final model has $m=8$ rep-GFBs blocks, the expansion coefficient $\beta$ is set to 16, the ghost rate $\alpha$ is set to 0.5, the number of channels $C$ before expanding is set to 28. The model was trained to minimize the L1 loss function on patches of size $64\times64$ pixels with a batch size of 32. Network parameters were optimized for 1000 epochs using the Adam optimizer with the initial learning rate of 1e--4 halved every 200 epochs. Random flips and rotations were used for data augmentation, quantization-aware training was applied to improve the accuracy of the resulting INT8 model.

\subsection{ICL}

Team ICL used the ABPN model, where $3\times3$ convolutions were treated as residual in residual re-parameterization blocks (RRRBs)~\cite{du2022fast} before the fine-tuning. Then, these blocks were fused using the re-parametrization trick, and the overall number of model parameters was decreased from 140K to 42.5K. The model was trained for 1000 epochs using the L1 loss function.

\subsection{Just A try}

Team Just A try also used the ABPN network, and decreased the number of feature channels from 28 to 12 for a better model efficiency. The network was first trained on the DIV2K dataset with a batch size of 16 on $64\times64$ pixel patches, and then fine-tuned on the DIV2K+Flicrk2K dataset with a batch size of 32 on $150\times150$ pixel patches.

\subsection{Bilibili AI}

\begin{figure*}[h!]
\centering
\includegraphics[width=1.0\linewidth]{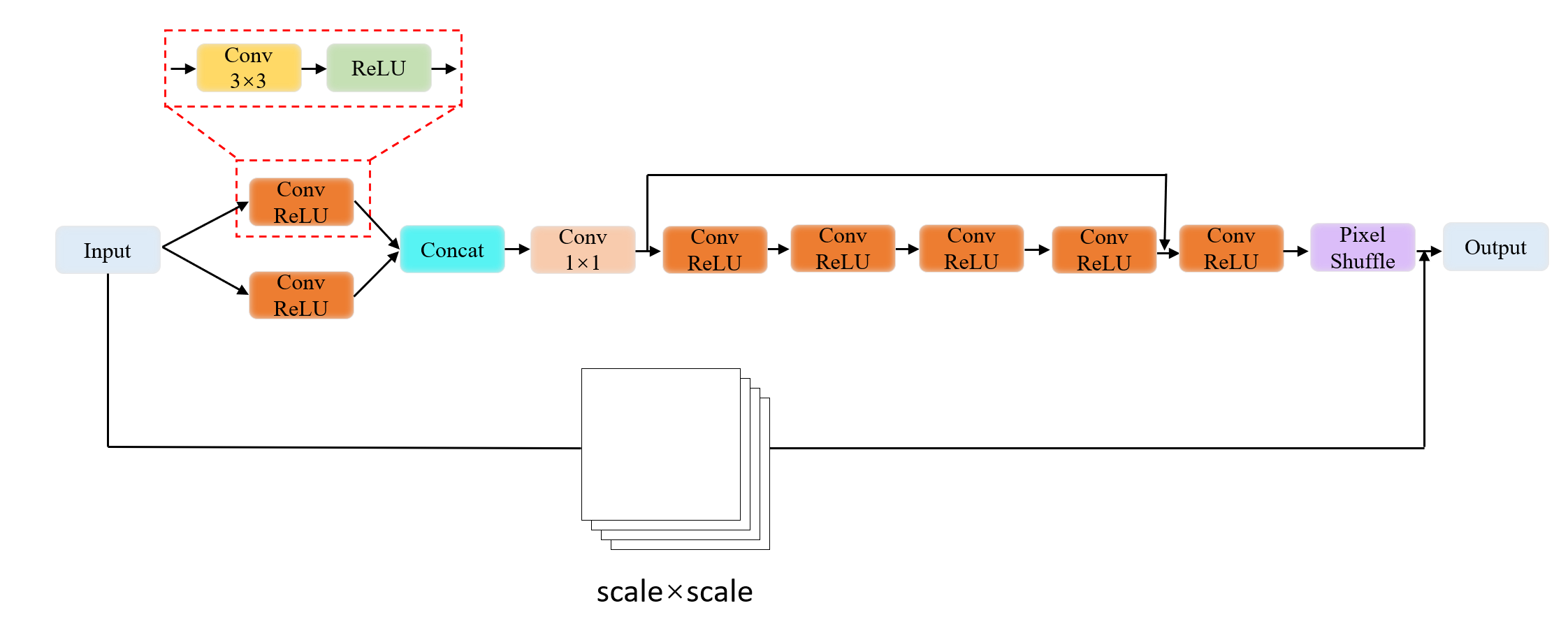}
\caption{\small{Model architecture proposed by team Bilibili AI.}}
\label{fig:Bilibili}
\end{figure*}

The architecture proposed Bilibili AI is shown in Fig.~\ref{fig:Bilibili}. The model first extracts image features through two identical convolutions followed by ReLU activations, concatenates them, and then performs an inter-channel information combination through a $1\times1$ convolution. Next, the obtained features are processed by several $3\times3$ convolutions with an additional skip connection. In the feature reconstruction layer, the authors obtain the final image through a pixel shuffle op and a skip connection adding the input image resized via pixel multiplication. The model was trained to minimize the L1 loss function on patches of size $48\times48$ pixels with a batch size of 16. Network parameters were optimized for 1080 epochs using the Adam optimizer with the initial learning rate of 1e--3 decreased by half every 120 epochs.

\subsection{MobileSR}

\begin{figure*}[h!]
\centering
\includegraphics[width=1.0\linewidth]{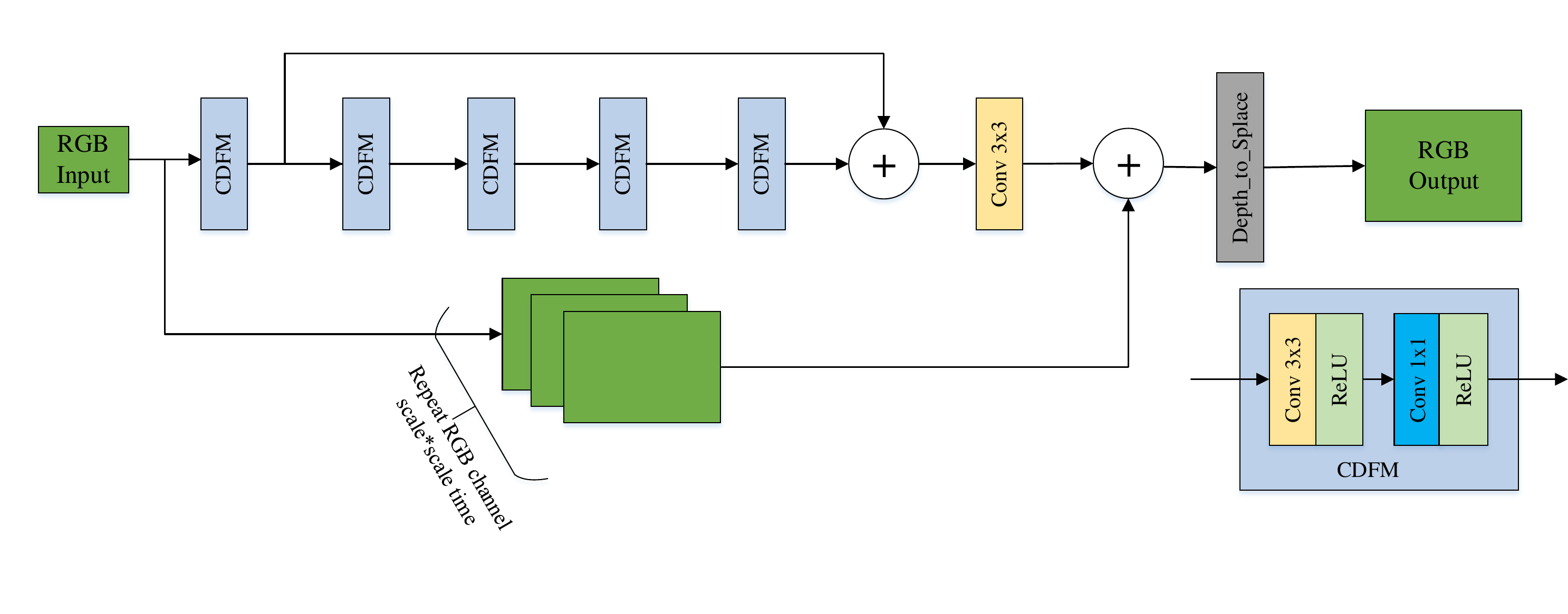}
\caption{\small{Model architecture proposed by team MobileSR.}}
\label{fig:MobileSR}
\end{figure*}

The model~\cite{gendy2022real} proposed by team MobileSR contains three blocks: the shallow feature extraction, the deep feature extraction, and the high-resolution image reconstruction blocks shown in Fig.~\ref{fig:MobileSR}. The shallow feature extraction block contains only one channel-and-deep-features-mixer (CDFM, 3$\times$3 conv followed by 3$\times$3 conv with ReLU activations after each one). After that, the deep feature extraction module contains 4 stacked CDFM blocks and one skip connection. The final module performing the image reconstruction is using one 3$\times$3 convolution, an anchor to transfer low-frequency information, and a depth-to-space layer used to produce the final HR image. The model was trained to minimize the L1 loss function on patches of size $64\times64$ pixels with a batch size of 32. Network parameters were optimized for 600 epochs using the Adam optimizer with the initial learning rate of 1e--3 halved every 120 epochs. Quantization-aware training was applied to improve the accuracy of the resulting INT8 model.

\subsection{Noahtcv}

\begin{figure*}[h!]
\centering
\includegraphics[width=0.6\linewidth]{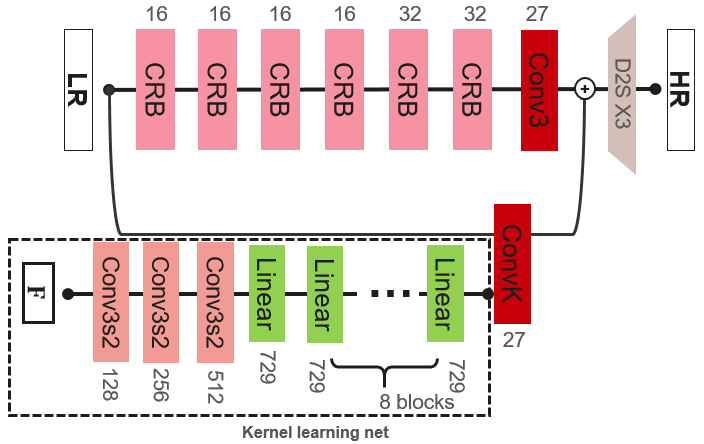}
\includegraphics[width=0.1\linewidth]{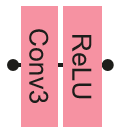}
\caption{\small{The overall model architecture and the structure of the CRB block (right) proposed by team Noahtcv.}}
\label{fig:Noahtcv}
\end{figure*}

Team noahtcv proposed the architecture that utilizes a pixel shuffling block and seven convolutions on the low resolution space to achieve faster processing times (Fig.~\ref{fig:Noahtcv}). The network is comprised of two branches connected by a global residual. The global skip connection learns upsampling kernels using a smaller network comprised of 3 convolutions with stride 2 and 9 linear blocks. Each layer of the kernel learning network (KLN) is passed to a ReLU activation unit. The input to the KLN network is image patches (optionally spatially frequencies too) from the low resolution image. Once the network converges, the top tensor of KLN is fed to the ConvK block generating 27 channels as weight variable to convert it to the inference network. 

The model was trained to minimize the L1 loss function on patches of size $64\times64$ pixels. Network parameters were optimized using the Adam optimizer with the initial learning rate of 25e--4 decreased by half every 150K iterations. The kernel learning subnet is trained once with the whole network. When the performance converges, the subnet is pruned and the output tensor is fed to the global skip ConvK. The global skip connection becomes a single convolutional layer at inference time.

\section{Additional Literature}

An overview of the past challenges on mobile-related tasks together with the proposed solutions can be found in the following papers:

\begin{itemize}
\item Image Super-Resolution:\, \cite{ignatov2018pirm,lugmayr2020ntire,cai2019ntire,timofte2018ntire}
\item Learned End-to-End ISP:\, \cite{ignatov2019aim,ignatov2020aim,ignatov2022microisp,ignatov2022pynetv2}
\item Perceptual Image Enhancement:\, \cite{ignatov2018pirm,ignatov2019ntire}
\item Bokeh Effect Rendering:\, \cite{ignatov2019aimBokeh,ignatov2020aimBokeh}
\item Image Denoising:\, \cite{abdelhamed2020ntire,abdelhamed2019ntire}
\end{itemize}

\section*{Acknowledgements}

We thank the sponsors of the Mobile AI and AIM 2022 workshops and challenges: Synaptics Inc., AI Witchlabs, MediaTek, Huawei, Reality Labs, OPPO, Raspberry Pi, ETH Z\"urich (Computer Vision Lab) and University of W\"urzburg (Computer Vision Lab).


\appendix
\section{Teams and Affiliations}
\label{sec:apd:team}

\bigskip

\subsection*{Mobile AI 2022 Team}
\noindent\textit{\textbf{Title: }}\\ Mobile AI 2022 Image Super-Resolution Challenge\\
\noindent\textit{\textbf{Members:}}\\ Andrey Ignatov$^{1,2}$ \textit{(andrey@vision.ee.ethz.ch)}, Radu Timofte$^{1,2,3}$  \textit{(radu.timofte @vision.ee.ethz.ch)}, Maurizio Denna$^4$ \textit{(maurizio.denna@synaptics.com)}, Abdel Younes$^4$ \textit{(abdel.younes@synaptics.com)}\\
\noindent\textit{\textbf{Affiliations: }}\\
$^1$ Computer Vision Lab, ETH Z\"urich, Switzerland\\
$^2$ AI Witchlabs, Switzerland\\
$^3$ University of W\"urzburg, Germany\\
$^4$ Synaptics Europe, Switzerland\\

\subsection*{Z6}
\noindent\textit{\textbf{Title:}}\\ Skip-Concatenated Image Super Resolution Network (SCSRN) for Mobile Devices\\
\noindent\textit{\textbf{Members:}}\\ \textit{Ganzorig Gankhuyag$^1$ (gnzrg25@gmail.com), Jingang Huh$^1$, Myeong Kyun Kim$^1$, Kihwan Yoon$^1$, Hyeon-Cheol Moon$^1$, Seungho Lee$^1$, Yoonsik Choe$^2$, Jinwoo Jeong$^1$, Sungjei Kim$^1$}\\
\noindent\textit{\textbf{Affiliations: }}\\
$^1$ Korea Electronics Technology Institute (KETI), South Korea\\
$^2$ Yonsei University, South Korea\\

\subsection*{TCLResearchEurope}
\noindent\textit{\textbf{Title:}}\\ RPQ~--- Extremely Efficient Super-Resolution Network Via Reparametrization, Pruning and Quantization\\
\noindent\textit{\textbf{Members:}}\\ \textit{Maciej Smyl (maciej.smyl@gmail.com)}, Tomasz Latkowski, Pawel Kubik, Michal Sokolski, Yujie Ma\\
\noindent\textit{\textbf{Affiliations: }}\\
TCL Research Europe, Poland\\

\subsection*{ECNUSR}
\noindent\textit{\textbf{Title:}}\\ PureConvSR: lightweight pure convolutional neural network with equivalent transformation\\
\noindent\textit{\textbf{Members:}}\\ \textit{Jiahao Chao (51215902006@stu.ecnu.edu.cn)}, Zhou Zhou, Hongfan Gao, Zhengfeng Yang, Zhenbing Zeng\\
\noindent\textit{\textbf{Affiliations: }}\\
East China Normal University, China\\

\subsection*{LCVG}
\noindent\textit{\textbf{Title:}}\\ HOPN: A Hardware Optimized Plain Network for Mobile Image Super-Resolution\\
\noindent\textit{\textbf{Members:}}\\ \textit{Zhengyang Zhuge (zyoung2333@gmail.com)}, Chenghua Li\\
\noindent\textit{\textbf{Affiliations: }}\\
Institute of Automation, Chinese Academy of Sciences, China\\

\subsection*{BOE-IOT-AIBD}
\noindent\textit{\textbf{Title:}}\\ Lightweight Quantization CNNNet for Mobile Image Super-Resolution\\
\noindent\textit{\textbf{Members:}}\\ \textit{Dan Zhu (zhudan@boe.com.cn)}, Mengdi Sun, Ran Duan, Yan Gao\\
\noindent\textit{\textbf{Affiliations: }}\\
BOE Technology Group Co., Ltd., China\\

\subsection*{NJUST}
\noindent\textit{\textbf{Title:}}\\ EMSRNet: An Efficient ConvNet for Real-time Image Super-resolution on Mobile Devices\\
\noindent\textit{\textbf{Members:}}\\ \textit{Lingshun Kong (konglingshun@njust.edu.cn)}, Long Sun, Xiang Li, Xingdong Zhang, Jiawei Zhang, Yaqi Wu, Jinshan Pan\\
\noindent\textit{\textbf{Affiliations: }}\\
Nanjing University of Science and Technology, China\\

\subsection*{Antins\_cv}
\noindent\textit{\textbf{Title:}}\\ Extremely Light-Weight Dual-Branch Network for Real Time Image Super Resolution\\
\noindent\textit{\textbf{Members:}}\\ \textit{Gaocheng Yu (yugaocheng.ygc@antgroup.com)}, Jin Zhang, Feng Zhang, Zhe Ma, Hongbin Wang\\
\noindent\textit{\textbf{Affiliations: }}\\
Ant Group, China\\

\subsection*{GenMedia Group}
\noindent\textit{\textbf{Title:}}\\ SkipSkip Video Super-Resolution \\
\noindent\textit{\textbf{Members: }}\\ \textit{Hojin Cho (jin@gengen.ai)}, Steve Kim\\
\noindent\textit{\textbf{Affiliations: }}\\
GenGenAI, South Korea\\

\subsection*{Vccip}
\noindent\textit{\textbf{Title:}}\\ Diverse Branch Re-Parameterizable Net for Mobile Image Super-Resolution\\
\noindent\textit{\textbf{Members:}}\\ \textit{Huaen Li (huaenli@mail.hfut.edu.cn)}, Yanbo Ma\\
\noindent\textit{\textbf{Affiliations: }}\\
Hefei University of Technology, China\\

\subsection*{MegSR}
\noindent\textit{\textbf{Title:}}\\ Fast Nearest Convolution for Real-Time Image Super-Resolution~\cite{luo2022fast}\\
\noindent\textit{\textbf{Members:}}\\ \textit{Ziwei Luo (ziwei.ro@gmail.com)}, Youwei Li, Lei Yu, Zhihong Wen, Qi Wu, Haoqiang Fan, Shuaicheng Liu\\
\noindent\textit{\textbf{Affiliations: }}\\
Megvii Technology, China\\
University of Electronic Science and Technology of China (UESTC), China\\

\subsection*{DoubleZ}
\noindent\textit{\textbf{Title:}}\\ Fast Image Super-Resolution Model\\
\noindent\textit{\textbf{Members:}}\\ \textit{Lize Zhang (lzzhang\_98@stu.xidian.edu.cn)}, Zhikai Zong\\
\noindent\textit{\textbf{Affiliations: }}\\
Xidian University, China\\
Qingdao Hi-image Technologies Co., Ltd., China\\
 
\subsection*{Jeremy Kwon}
\noindent\textit{\textbf{Title:}}\\ S2R2: Salus Super-Resolution Research\\
\noindent\textit{\textbf{Members:}}\\ \textit{Jeremy Kwon (alan.jaeger0@gmail.com)}\\
\noindent\textit{\textbf{Affiliations: }}\\
None, South Korea\\

\subsection*{Lab216}
\noindent\textit{\textbf{Title:}}\\ Lightweight Asymmetric Super-Resolution Network with Contrastive Quantized-aware Training\\
\noindent\textit{\textbf{Members:}}\\ \textit{Junxi Zhang (sissie\_zhang@whu.edu.cn)}, Mengyuan Li, Nianxiang Fu, Guanchen Ding, Han Zhu, Zhenzhong Chen\\
\noindent\textit{\textbf{Affiliations: }}\\
Wuhan University, China\\

\subsection*{TOVB}
\noindent\textit{\textbf{Title:}}\\ RBPN: Repconv-based Plain Net for Mobile Image Super-Resolution\\
\noindent\textit{\textbf{Members:}}\\ \textit{Gen Li (leegeun@yonsei.ac.kr)}, Yuanfan Zhang, Lei Sun\\
\noindent\textit{\textbf{Affiliations: }}\\
None, China\\

\subsection*{Samsung Research}
\noindent\textit{\textbf{Title:}}\\ ABPN++: Anchor-based Plain Net for Mobile Image Super-Resolution with Pre-training\\
\noindent\textit{\textbf{Members:}}\\ \textit{Dafeng Zhang (dfeng.zhang@samsung.com)}\\
\noindent\textit{\textbf{Affiliations: }}\\
Samsung Research, China\\

\subsection*{Rtsisr2022}
\noindent\textit{\textbf{Title:}}\\ Re-parameterized Anchor-based Plain Network\\
\noindent\textit{\textbf{Members:}}\\ \textit{Neo Yang (296859095@qq.com)}, Fitz Liu, Jerry Zhao\\
\noindent\textit{\textbf{Affiliations: }}\\
None, China\\

\subsection*{Aselsan Research}
\noindent\textit{\textbf{Title:}}\\XCAT - Lightweight Single Image Super Resolution Network with Cross Concatenated Heterogeneous Group Convolutions~\cite{ayazoglu2022xcat}\\
\noindent\textit{\textbf{Members:}}\\ \textit{Mustafa Ayazoglu (mayazoglu@aselsan.com.tr)}, Bahri Batuhan Bilecen\\
\noindent\textit{\textbf{Affiliations: }}\\
Aselsan Corporation, Turkey\\

\subsection*{Klab\_SR}
\noindent\textit{\textbf{Title:}}\\ Deeper and narrower SR Model\\
\noindent\textit{\textbf{Members:}}\\ \textit{Shota Hirose (syouta.hrs@akane.waseda.jp)}, Kasidis Arunruangsirilert, Luo Ao\\
\noindent\textit{\textbf{Affiliations: }}\\
Waseda University, Japan\\

\subsection*{TCL Research HK}
\noindent\textit{\textbf{Title:}}\\ Anchor-collapsed Super Resolution for Mobile Devices\\
\noindent\textit{\textbf{Members:}}\\ \textit{Ho Chun Leung (hcleung@tcl.com)}, Andrew Wei, Jie Liu, Qiang Liu, Dahai Yu\\
\noindent\textit{\textbf{Affiliations: }}\\
TCL Corporate Research, China\\

\subsection*{RepGSR}
\noindent\textit{\textbf{Title:}}\\ Super Lightweight Super-resolution Based on Ghost Features with Re-parameterization\\
\noindent\textit{\textbf{Members:}}\\ \textit{Ao Li (liao@cqu.edu.cn)}, Lei Luo, Ce Zhu\\
\noindent\textit{\textbf{Affiliations: }}\\
University of Electronic Science and Technology, China\\

\subsection*{ICL}
\noindent\textit{\textbf{Title:}}\\ Neural Network Quantization With Reparametrization And Kernel Range Regularization\\
\noindent\textit{\textbf{Members:}}\\ \textit{Seongmin Hong (smhongok@snu.ac.kr)}, Dongwon Park, Joonhee Lee, Byeong Hyun Lee, Seunggyu Lee, Se Young Chun\\
\noindent\textit{\textbf{Affiliations: }}\\
Intelligent Computational Imaging Lab, Seoul National University, South Korea\\

\subsection*{Just A try}
\noindent\textit{\textbf{Title:}}\\ Plain net for realtime Super-resolution\\
\noindent\textit{\textbf{Members:}}\\ \textit{Ruiyuan He (164643209@qq.com)}, Xuhao Jiang\\
\noindent\textit{\textbf{Affiliations: }}\\
None, China\\

\subsection*{Bilibili AI}
\noindent\textit{\textbf{Title:}}\\ A Robust Anchor-based network for Mobile Super-Resolution\\
\noindent\textit{\textbf{Members:}}\\ \textit{Haihang Ruan (hhruan@mail.sim.ac.cn)}, Xinjian Zhang, Jing Liu\\
\noindent\textit{\textbf{Affiliations: }}\\
Bilibili Inc., China\\

\subsection*{MobileSR}
\noindent\textit{\textbf{Title:}}\\ Real-Time Channel Mixing Net for Mobile Image Super-Resolution~\cite{gendy2022real}\\
\noindent\textit{\textbf{Members:}}\\ \textit{Garas Gendy$^1$ (garasgaras@yahoo.com)}, Nabil Sabor$^2$, Jingchao Hou$^1$, Guanghui He$^1$\\
\noindent\textit{\textbf{Affiliations: }}\\
$^1$ Shanghai Jiao Tong University, China\\
$^2$ Assiut University, Egypt \\

{\small
\bibliographystyle{splncs04}
\bibliography{egbib}

\begin{thebibliography}{10}
\providecommand{\url}[1]{\texttt{#1}}
\providecommand{\urlprefix}{URL }
\providecommand{\doi}[1]{https://doi.org/#1}

\bibitem{abdelhamed2020ntire}
Abdelhamed, A., Afifi, M., Timofte, R., Brown, M.S.: Ntire 2020 challenge on
  real image denoising: Dataset, methods and results. In: Proceedings of the
  IEEE/CVF Conference on Computer Vision and Pattern Recognition Workshops. pp.
  496--497 (2020)

\bibitem{abdelhamed2019ntire}
Abdelhamed, A., Timofte, R., Brown, M.S.: Ntire 2019 challenge on real image
  denoising: Methods and results. In: Proceedings of the IEEE/CVF Conference on
  Computer Vision and Pattern Recognition Workshops. pp.~0--0 (2019)

\bibitem{agustsson2017ntire}
Agustsson, E., Timofte, R.: Ntire 2017 challenge on single image
  super-resolution: Dataset and study. In: Proceedings of the IEEE Conference
  on Computer Vision and Pattern Recognition Workshops. pp. 126--135 (2017)

\bibitem{anwar2017structured}
Anwar, S., Hwang, K., Sung, W.: Structured pruning of deep convolutional neural
  networks. ACM Journal on Emerging Technologies in Computing Systems (JETC)
  \textbf{13}(3),  1--18 (2017)

\bibitem{anwar2016compact}
Anwar, S., Sung, W.: Compact deep convolutional neural networks with coarse
  pruning. arXiv preprint arXiv:1610.09639  (2016)

\bibitem{ayazoglu2021extremely}
Ayazoglu, M.: Extremely lightweight quantization robust real-time single-image
  super resolution for mobile devices. In: Proceedings of the IEEE/CVF
  Conference on Computer Vision and Pattern Recognition Workshops. pp.~0--0
  (2021)

\bibitem{ayazoglu2022xcat}
Ayazoglu, M., Bilecen, B.B.: {XCAT} – lightweight quantized single image
  super-resolution using heterogeneous group convolutions and cross
  concatenation. In: Proceedings of the European Conference on Computer Vision
  (ECCV) Workshops (2022)

\bibitem{bhardwaj2022collapsible}
Bhardwaj, K., Milosavljevic, M., O'Neil, L., Gope, D., Matas, R., Chalfin, A.,
  Suda, N., Meng, L., Loh, D.: Collapsible linear blocks for super-efficient
  super resolution. Proceedings of Machine Learning and Systems  \textbf{4},
  529--547 (2022)

\bibitem{cai2019ntire}
Cai, J., Gu, S., Timofte, R., Zhang, L.: Ntire 2019 challenge on real image
  super-resolution: Methods and results. In: Proceedings of the IEEE/CVF
  Conference on Computer Vision and Pattern Recognition Workshops. pp.~0--0
  (2019)

\bibitem{cai2020zeroq}
Cai, Y., Yao, Z., Dong, Z., Gholami, A., Mahoney, M.W., Keutzer, K.: Zeroq: A
  novel zero shot quantization framework. In: Proceedings of the IEEE/CVF
  Conference on Computer Vision and Pattern Recognition. pp. 13169--13178
  (2020)

\bibitem{chiang2020deploying}
Chiang, C.M., Tseng, Y., Xu, Y.S., Kuo, H.K., Tsai, Y.M., Chen, G.Y., Tan,
  K.S., Wang, W.T., Lin, Y.C., Tseng, S.Y.R., et~al.: Deploying image
  deblurring across mobile devices: A perspective of quality and latency. In:
  Proceedings of the IEEE/CVF Conference on Computer Vision and Pattern
  Recognition Workshops. pp. 502--503 (2020)

\bibitem{conde2022aim}
Conde, M.V., Timofte, R., et~al.: {R}eversed {I}mage {S}ignal {P}rocessing and
  {RAW} {R}econstruction. {AIM} 2022 {C}hallenge {R}eport. In: Proceedings of
  the European Conference on Computer Vision (ECCV) Workshops (2022)

\bibitem{ding2021diverse}
Ding, X., Zhang, X., Han, J., Ding, G.: Diverse branch block: Building a
  convolution as an inception-like unit. In: Proceedings of the IEEE/CVF
  Conference on Computer Vision and Pattern Recognition. pp. 10886--10895
  (2021)

\bibitem{ding2021repvgg}
Ding, X., Zhang, X., Ma, N., Han, J., Ding, G., Sun, J.: Repvgg: Making
  vgg-style convnets great again. In: Proceedings of the IEEE/CVF Conference on
  Computer Vision and Pattern Recognition. pp. 13733--13742 (2021)

\bibitem{dong2014learning}
Dong, C., Loy, C.C., He, K., Tang, X.: Learning a deep convolutional network
  for image super-resolution. In: European conference on computer vision. pp.
  184--199. Springer (2014)

\bibitem{dong2015image}
Dong, C., Loy, C.C., He, K., Tang, X.: Image super-resolution using deep
  convolutional networks. IEEE transactions on pattern analysis and machine
  intelligence  \textbf{38}(2),  295--307 (2015)

\bibitem{dozat2016incorporating}
Dozat, T.: Incorporating nesterov momentum into adam  (2016)

\bibitem{du2022fast}
Du, Z., Liu, D., Liu, J., Tang, J., Wu, G., Fu, L.: Fast and memory-efficient
  network towards efficient image super-resolution. In: Proceedings of the
  IEEE/CVF Conference on Computer Vision and Pattern Recognition. pp. 853--862
  (2022)

\bibitem{du2021anchor}
Du, Z., Liu, J., Tang, J., Wu, G.: Anchor-based plain net for mobile image
  super-resolution. In: Proceedings of the IEEE/CVF Conference on Computer
  Vision and Pattern Recognition Workshops. pp.~0--0 (2021)

\bibitem{freeman2002example}
Freeman, W.T., Jones, T.R., Pasztor, E.C.: Example-based super-resolution. IEEE
  Computer graphics and Applications  \textbf{22}(2),  56--65 (2002)

\bibitem{gendy2022real}
Gendy, G., nabil sabor, Hou, J., He, G.: Real-time channel mixing net for
  mobile image super-resolution. In: Proceedings of the European Conference on
  Computer Vision (ECCV) Workshops (2022)

\bibitem{howard2019searching}
Howard, A., Sandler, M., Chu, G., Chen, L.C., Chen, B., Tan, M., Wang, W., Zhu,
  Y., Pang, R., Vasudevan, V., et~al.: Searching for mobilenetv3. In:
  Proceedings of the IEEE/CVF International Conference on Computer Vision. pp.
  1314--1324 (2019)

\bibitem{huang2015single}
Huang, J.B., Singh, A., Ahuja, N.: Single image super-resolution from
  transformed self-exemplars. In: Proceedings of the IEEE conference on
  computer vision and pattern recognition. pp. 5197--5206 (2015)

\bibitem{ignatov2021fastDenoising}
Ignatov, A., Byeoung-su, K., Timofte, R.: Fast camera image denoising on mobile
  gpus with deep learning, mobile ai 2021 challenge: Report. In: Proceedings of
  the IEEE/CVF Conference on Computer Vision and Pattern Recognition Workshops.
  pp.~0--0 (2021)

\bibitem{ignatov2021learned}
Ignatov, A., Chiang, J., Kuo, H.K., Sycheva, A., Timofte, R.: Learned
  smartphone isp on mobile npus with deep learning, mobile ai 2021 challenge:
  Report. In: Proceedings of the IEEE/CVF Conference on Computer Vision and
  Pattern Recognition Workshops. pp.~0--0 (2021)

\bibitem{ignatov2017dslr}
Ignatov, A., Kobyshev, N., Timofte, R., Vanhoey, K., Van~Gool, L.: Dslr-quality
  photos on mobile devices with deep convolutional networks. In: Proceedings of
  the IEEE International Conference on Computer Vision. pp. 3277--3285 (2017)

\bibitem{ignatov2018wespe}
Ignatov, A., Kobyshev, N., Timofte, R., Vanhoey, K., Van~Gool, L.: Wespe:
  weakly supervised photo enhancer for digital cameras. In: Proceedings of the
  IEEE Conference on Computer Vision and Pattern Recognition Workshops. pp.
  691--700 (2018)

\bibitem{ignatov2021fastDepth}
Ignatov, A., Malivenko, G., Plowman, D., Shukla, S., Timofte, R.: Fast and
  accurate single-image depth estimation on mobile devices, mobile ai 2021
  challenge: Report. In: Proceedings of the IEEE/CVF Conference on Computer
  Vision and Pattern Recognition Workshops. pp.~0--0 (2021)

\bibitem{ignatov2021fastSceneDetection}
Ignatov, A., Malivenko, G., Timofte, R.: Fast and accurate quantized camera
  scene detection on smartphones, mobile ai 2021 challenge: Report. In:
  Proceedings of the IEEE/CVF Conference on Computer Vision and Pattern
  Recognition Workshops. pp.~0--0 (2021)

\bibitem{ignatov2022pynetv2}
Ignatov, A., Malivenko, G., Timofte, R., Tseng, Y., Xu, Y.S., Yu, P.H., Chiang,
  C.M., Kuo, H.K., Chen, M.H., Cheng, C.M., Van~Gool, L.: Pynet-v2 mobile:
  Efficient on-device photo processing with neural networks. In: 2021 26th
  International Conference on Pattern Recognition (ICPR). IEEE (2022)

\bibitem{ignatov2022maidepth}
Ignatov, A., Malivenko, G., Timofte, R., et~al.: Efficient single-image depth
  estimation on mobile devices, mobile ai \& aim 2022 challenge: Report. In:
  European Conference on Computer Vision (2022)

\bibitem{ignatov2020rendering}
Ignatov, A., Patel, J., Timofte, R.: Rendering natural camera bokeh effect with
  deep learning. In: Proceedings of the IEEE/CVF Conference on Computer Vision
  and Pattern Recognition Workshops. pp. 418--419 (2020)

\bibitem{ignatov2019aimBokeh}
Ignatov, A., Patel, J., Timofte, R., Zheng, B., Ye, X., Huang, L., Tian, X.,
  Dutta, S., Purohit, K., Kandula, P., et~al.: Aim 2019 challenge on bokeh
  effect synthesis: Methods and results. In: 2019 IEEE/CVF International
  Conference on Computer Vision Workshop (ICCVW). pp. 3591--3598. IEEE (2019)

\bibitem{romero2021real}
Ignatov, A., Romero, A., Kim, H., Timofte, R.: Real-time video super-resolution
  on smartphones with deep learning, mobile ai 2021 challenge: Report. In:
  Proceedings of the IEEE/CVF Conference on Computer Vision and Pattern
  Recognition Workshops. pp.~0--0 (2021)

\bibitem{ignatov2022microisp}
Ignatov, A., Sycheva, A., Timofte, R., Tseng, Y., Xu, Y.S., Yu, P.H., Chiang,
  C.M., Kuo, H.K., Chen, M.H., Cheng, C.M., Van~Gool, L.: Microisp: Processing
  32mp photos on mobile devices with deep learning. In: European Conference on
  Computer Vision (2022)

\bibitem{ignatov2019ntire}
Ignatov, A., Timofte, R.: Ntire 2019 challenge on image enhancement: Methods
  and results. In: Proceedings of the IEEE/CVF Conference on Computer Vision
  and Pattern Recognition Workshops. pp.~0--0 (2019)

\bibitem{ignatov2022maivideosr}
Ignatov, A., Timofte, R., Chiang, C.M., Kuo, H.K., Xu, Y.S., Lee, M.Y., Lu, A.,
  Cheng, C.M., Chen, C.C., Yong, J.Y., Shuai, H.H., Cheng, W.H., et~al.: Power
  efficient video super-resolution on mobile npus with deep learning, mobile ai
  \& aim 2022 challenge: Report. In: European Conference on Computer Vision
  (2022)

\bibitem{ignatov2018ai}
Ignatov, A., Timofte, R., Chou, W., Wang, K., Wu, M., Hartley, T., Van~Gool,
  L.: Ai benchmark: Running deep neural networks on android smartphones. In:
  Proceedings of the European Conference on Computer Vision (ECCV) Workshops.
  pp.~0--0 (2018)

\bibitem{ignatov2021real}
Ignatov, A., Timofte, R., Denna, M., Younes, A.: Real-time quantized image
  super-resolution on mobile npus, mobile ai 2021 challenge: Report. In:
  Proceedings of the IEEE/CVF Conference on Computer Vision and Pattern
  Recognition Workshops. pp.~0--0 (2021)

\bibitem{ignatov2019aim}
Ignatov, A., Timofte, R., Ko, S.J., Kim, S.W., Uhm, K.H., Ji, S.W., Cho, S.J.,
  Hong, J.P., Mei, K., Li, J., et~al.: Aim 2019 challenge on raw to rgb
  mapping: Methods and results. In: 2019 IEEE/CVF International Conference on
  Computer Vision Workshop (ICCVW). pp. 3584--3590. IEEE (2019)

\bibitem{ignatov2019ai}
Ignatov, A., Timofte, R., Kulik, A., Yang, S., Wang, K., Baum, F., Wu, M., Xu,
  L., Van~Gool, L.: Ai benchmark: All about deep learning on smartphones in
  2019. In: 2019 IEEE/CVF International Conference on Computer Vision Workshop
  (ICCVW). pp. 3617--3635. IEEE (2019)

\bibitem{ignatov2020aimBokeh}
Ignatov, A., Timofte, R., Qian, M., Qiao, C., Lin, J., Guo, Z., Li, C., Leng,
  C., Cheng, J., Peng, J., et~al.: Aim 2020 challenge on rendering realistic
  bokeh. In: European Conference on Computer Vision. pp. 213--228. Springer
  (2020)

\bibitem{ignatov2018pirm}
Ignatov, A., Timofte, R., Van~Vu, T., Minh~Luu, T., X~Pham, T., Van~Nguyen, C.,
  Kim, Y., Choi, J.S., Kim, M., Huang, J., et~al.: Pirm challenge on perceptual
  image enhancement on smartphones: Report. In: Proceedings of the European
  Conference on Computer Vision (ECCV) Workshops. pp.~0--0 (2018)

\bibitem{ignatov2020aim}
Ignatov, A., Timofte, R., Zhang, Z., Liu, M., Wang, H., Zuo, W., Zhang, J.,
  Zhang, R., Peng, Z., Ren, S., et~al.: Aim 2020 challenge on learned image
  signal processing pipeline. arXiv preprint arXiv:2011.04994  (2020)

\bibitem{ignatov2022maiisp}
Ignatov, A., Timofte, R., et~al.: Learned smartphone isp on mobile gpus with
  deep learning, mobile ai \& aim 2022 challenge: Report. In: European
  Conference on Computer Vision (2022)

\bibitem{ignatov2022maibokeh}
Ignatov, A., Timofte, R., et~al.: Realistic bokeh effect rendering on mobile
  gpus, mobile ai \& aim 2022 challenge: Report (2022)

\bibitem{ignatov2020replacing}
Ignatov, A., Van~Gool, L., Timofte, R.: Replacing mobile camera isp with a
  single deep learning model. In: Proceedings of the IEEE/CVF Conference on
  Computer Vision and Pattern Recognition Workshops. pp. 536--537 (2020)

\bibitem{ignatov2020controlling}
Ignatov, D., Ignatov, A.: Controlling information capacity of binary neural
  network. Pattern Recognition Letters  \textbf{138},  276--281 (2020)

\bibitem{SynapticsSoC2021}
Inc., S.: \url{https://www.synaptics.com/technology/edge-computing}

\bibitem{irani1991improving}
Irani, M., Peleg, S.: Improving resolution by image registration. CVGIP:
  Graphical models and image processing  \textbf{53}(3),  231--239 (1991)

\bibitem{jacob2018quantization}
Jacob, B., Kligys, S., Chen, B., Zhu, M., Tang, M., Howard, A., Adam, H.,
  Kalenichenko, D.: Quantization and training of neural networks for efficient
  integer-arithmetic-only inference. In: Proceedings of the IEEE Conference on
  Computer Vision and Pattern Recognition. pp. 2704--2713 (2018)

\bibitem{jain2019trained}
Jain, S.R., Gural, A., Wu, M., Dick, C.H.: Trained quantization thresholds for
  accurate and efficient fixed-point inference of deep neural networks. arXiv
  preprint arXiv:1903.08066  (2019)

\bibitem{kim2016accurate}
Kim, J., Lee, J.K., Lee, K.M.: Accurate image super-resolution using very deep
  convolutional networks. In: Proceedings of the IEEE conference on computer
  vision and pattern recognition. pp. 1646--1654 (2016)

\bibitem{kinli2022aim}
Kinli, F.O., Mentes, S., Ozcan, B., Kirac, F., Timofte, R., et~al.: Aim 2022
  challenge on instagram filter removal: Methods and results. In: Proceedings
  of the European Conference on Computer Vision (ECCV) Workshops (2022)

\bibitem{kong2022residual}
Kong, F., Li, M., Liu, S., Liu, D., He, J., Bai, Y., Chen, F., Fu, L.: Residual
  local feature network for efficient super-resolution. In: Proceedings of the
  IEEE/CVF Conference on Computer Vision and Pattern Recognition. pp. 766--776
  (2022)

\bibitem{krizhevsky2012imagenet}
Krizhevsky, A., Sutskever, I., Hinton, G.E.: Imagenet classification with deep
  convolutional neural networks. Advances in neural information processing
  systems  \textbf{25} (2012)

\bibitem{li2019learning}
Li, Y., Gu, S., Gool, L.V., Timofte, R.: Learning filter basis for
  convolutional neural network compression. In: Proceedings of the IEEE/CVF
  International Conference on Computer Vision. pp. 5623--5632 (2019)

\bibitem{li2022ntire}
Li, Y., Zhang, K., Timofte, R., Van~Gool, L., Kong, F., Li, M., Liu, S., Du,
  Z., Liu, D., Zhou, C., et~al.: Ntire 2022 challenge on efficient
  super-resolution: Methods and results. In: Proceedings of the IEEE/CVF
  Conference on Computer Vision and Pattern Recognition. pp. 1062--1102 (2022)

\bibitem{lim2017enhanced}
Lim, B., Son, S., Kim, H., Nah, S., Mu~Lee, K.: Enhanced deep residual networks
  for single image super-resolution. In: Proceedings of the IEEE conference on
  computer vision and pattern recognition workshops. pp. 136--144 (2017)

\bibitem{liu2019metapruning}
Liu, Z., Mu, H., Zhang, X., Guo, Z., Yang, X., Cheng, K.T., Sun, J.:
  Metapruning: Meta learning for automatic neural network channel pruning. In:
  Proceedings of the IEEE/CVF International Conference on Computer Vision. pp.
  3296--3305 (2019)

\bibitem{liu2018bi}
Liu, Z., Wu, B., Luo, W., Yang, X., Liu, W., Cheng, K.T.: Bi-real net:
  Enhancing the performance of 1-bit cnns with improved representational
  capability and advanced training algorithm. In: Proceedings of the European
  conference on computer vision (ECCV). pp. 722--737 (2018)

\bibitem{loshchilov2016sgdr}
Loshchilov, I., Hutter, F.: Sgdr: Stochastic gradient descent with warm
  restarts. arXiv preprint arXiv:1608.03983  (2016)

\bibitem{lugmayr2020ntire}
Lugmayr, A., Danelljan, M., Timofte, R.: Ntire 2020 challenge on real-world
  image super-resolution: Methods and results. In: Proceedings of the IEEE/CVF
  Conference on Computer Vision and Pattern Recognition Workshops. pp. 494--495
  (2020)

\bibitem{luo2022fast}
Luo, Z., Li, Y., lei yu, Wu, Q., wen zhihong, Fan, H., Liu, S.: Fast nearest
  convolution for real-time efficient image super-resolution. In: Proceedings
  of the European Conference on Computer Vision (ECCV) Workshops (2022)

\bibitem{obukhov2020t}
Obukhov, A., Rakhuba, M., Georgoulis, S., Kanakis, M., Dai, D., Van~Gool, L.:
  T-basis: a compact representation for neural networks. In: International
  Conference on Machine Learning. pp. 7392--7404. PMLR (2020)

\bibitem{park2003super}
Park, S.C., Park, M.K., Kang, M.G.: Super-resolution image reconstruction: a
  technical overview. IEEE signal processing magazine  \textbf{20}(3),  21--36
  (2003)

\bibitem{sandler2018mobilenetv2}
Sandler, M., Howard, A., Zhu, M., Zhmoginov, A., Chen, L.C.: Mobilenetv2:
  Inverted residuals and linear bottlenecks. In: Proceedings of the IEEE
  conference on computer vision and pattern recognition. pp. 4510--4520 (2018)

\bibitem{tan2019mnasnet}
Tan, M., Chen, B., Pang, R., Vasudevan, V., Sandler, M., Howard, A., Le, Q.V.:
  Mnasnet: Platform-aware neural architecture search for mobile. In:
  Proceedings of the IEEE/CVF Conference on Computer Vision and Pattern
  Recognition. pp. 2820--2828 (2019)

\bibitem{TensorFlowLite2021}
TensorFlow-Lite: \url{https://www.tensorflow.org/lite}

\bibitem{timofte2017ntire}
Timofte, R., Agustsson, E., Van~Gool, L., Yang, M.H., Zhang, L.: Ntire 2017
  challenge on single image super-resolution: Methods and results. In:
  Proceedings of the IEEE conference on computer vision and pattern recognition
  workshops. pp. 114--125 (2017)

\bibitem{timofte2013anchored}
Timofte, R., De~Smet, V., Van~Gool, L.: Anchored neighborhood regression for
  fast example-based super-resolution. In: Proceedings of the IEEE
  international conference on computer vision. pp. 1920--1927 (2013)

\bibitem{timofte2014a+}
Timofte, R., De~Smet, V., Van~Gool, L.: A+: Adjusted anchored neighborhood
  regression for fast super-resolution. In: Asian conference on computer
  vision. pp. 111--126. Springer (2014)

\bibitem{timofte2018ntire}
Timofte, R., Gu, S., Wu, J., Van~Gool, L.: Ntire 2018 challenge on single image
  super-resolution: Methods and results. In: Proceedings of the IEEE conference
  on computer vision and pattern recognition workshops. pp. 852--863 (2018)

\bibitem{timofte2016seven}
Timofte, R., Rothe, R., Van~Gool, L.: Seven ways to improve example-based
  single image super resolution. In: Proceedings of the IEEE Conference on
  Computer Vision and Pattern Recognition. pp. 1865--1873 (2016)

\bibitem{uhlich2019mixed}
Uhlich, S., Mauch, L., Cardinaux, F., Yoshiyama, K., Garcia, J.A., Tiedemann,
  S., Kemp, T., Nakamura, A.: Mixed precision dnns: All you need is a good
  parametrization. arXiv preprint arXiv:1905.11452  (2019)

\bibitem{wan2020fbnetv2}
Wan, A., Dai, X., Zhang, P., He, Z., Tian, Y., Xie, S., Wu, B., Yu, M., Xu, T.,
  Chen, K., et~al.: Fbnetv2: Differentiable neural architecture search for
  spatial and channel dimensions. In: Proceedings of the IEEE/CVF Conference on
  Computer Vision and Pattern Recognition. pp. 12965--12974 (2020)

\bibitem{wang2021fully}
Wang, H., Chen, P., Zhuang, B., Shen, C.: Fully quantized image
  super-resolution networks. In: Proceedings of the 29th ACM International
  Conference on Multimedia. pp. 639--647 (2021)

\bibitem{wang2021towards}
Wang, Y., Lin, S., Qu, Y., Wu, H., Zhang, Z., Xie, Y., Yao, A.: Towards compact
  single image super-resolution via contrastive self-distillation. arXiv
  preprint arXiv:2105.11683  (2021)

\bibitem{wu2019fbnet}
Wu, B., Dai, X., Zhang, P., Wang, Y., Sun, F., Wu, Y., Tian, Y., Vajda, P.,
  Jia, Y., Keutzer, K.: Fbnet: Hardware-aware efficient convnet design via
  differentiable neural architecture search. In: Proceedings of the IEEE/CVF
  Conference on Computer Vision and Pattern Recognition. pp. 10734--10742
  (2019)

\bibitem{yang2013fast}
Yang, C.Y., Yang, M.H.: Fast direct super-resolution by simple functions. In:
  Proceedings of the IEEE international conference on computer vision. pp.
  561--568 (2013)

\bibitem{yang2008image}
Yang, J., Wright, J., Huang, T., Ma, Y.: Image super-resolution as sparse
  representation of raw image patches. In: 2008 IEEE conference on computer
  vision and pattern recognition. pp.~1--8. IEEE (2008)

\bibitem{yang2010image}
Yang, J., Wright, J., Huang, T.S., Ma, Y.: Image super-resolution via sparse
  representation. IEEE transactions on image processing  \textbf{19}(11),
  2861--2873 (2010)

\bibitem{yang2019quantization}
Yang, J., Shen, X., Xing, J., Tian, X., Li, H., Deng, B., Huang, J., Hua, X.s.:
  Quantization networks. In: Proceedings of the IEEE/CVF Conference on Computer
  Vision and Pattern Recognition. pp. 7308--7316 (2019)

\bibitem{yang2022aim}
Yang, R., Timofte, R., et~al.: Aim 2022 challenge on super-resolution of
  compressed image and video: Dataset, methods and results. In: Proceedings of
  the European Conference on Computer Vision (ECCV) Workshops (2022)

\bibitem{zagoruyko2017diracnets}
Zagoruyko, S., Komodakis, N.: Diracnets: Training very deep neural networks
  without skip-connections. arXiv preprint arXiv:1706.00388  (2017)

\bibitem{zhang2020ntire}
Zhang, K., Gu, S., Timofte, R.: Ntire 2020 challenge on perceptual extreme
  super-resolution: Methods and results. In: Proceedings of the IEEE/CVF
  Conference on Computer Vision and Pattern Recognition Workshops. pp. 492--493
  (2020)

\bibitem{zhuang2020adabelief}
Zhuang, J., Tang, T., Ding, Y., Tatikonda, S.C., Dvornek, N., Papademetris, X.,
  Duncan, J.: Adabelief optimizer: Adapting stepsizes by the belief in observed
  gradients. Advances in neural information processing systems  \textbf{33},
  18795--18806 (2020)

\end{thebibliography}
}

\end{document}